\documentclass[epj]{svjour2}
\usepackage{graphics}
\usepackage{color}

\usepackage{graphicx}
\usepackage{epsf} 
\usepackage{slashed}
\usepackage{amsmath}
\usepackage{amssymb}

\def\be{\begin{equation}}
\def\ee{\end{equation}}
\def\ba{\begin{eqnarray}}
\def\ea{\end{eqnarray}}

\def\sfrac#1#2{{\textstyle \frac{#1}{#2}}}

\usepackage{url}

\begin{document}
\title{Combined parametrization of $G_{En}$
and $\gamma^\ast N \to \Delta(1232)$ \\ quadrupole form factors}
\author{G.~Ramalho  
}                     
%
%
\institute{Laborat\'orio de F\'{i}sica Te\'orica e Computacional -- LFTC, 
Universidade Cruzeiro do Sul, 01506-000, S\~ao Paulo, SP, Brazil}
\date{Received: date / Revised version: date}
%
\abstract{
Models based on $SU(6)$ symmetry breaking and large $N_c$ limit
provide relations between the pion cloud contributions to the 
$\gamma^\ast N \to \Delta(1232)$ quadrupole form factors,
electric ($G_E$) and Coulomb ($G_C$),
and the neutron electric form factor $G_{En}$,
suggesting that those form factors are dominated 
by the same physical processes. 
Those relations are improved in order to satisfy 
a fundamental constraint between the 
electric and Coulomb quadrupole form factors 
in the long wavelength limit, when the photon three-momentum vanishes 
(Siegert's theorem). 
Inspired by those relations, we study alternative parametrizations 
for the neutron electric form factor.
The parameters of the new form are then determined 
by a combined fit to the $G_{En}$ and the 
$\gamma^\ast N \to \Delta(1232)$ quadrupole form factor data.
We obtain a very good description of the $G_E$ and $G_C$   
data when we combine the pion cloud contributions 
with small valence quark contributions to the 
$\gamma^\ast N \to \Delta(1232)$ quadrupole form factors. 
The best description of the data is obtained 
when the second momentum of $G_{En}$ is $r_n^4 \simeq -0.4$ fm$^4$.
We conclude that the square radii associated with $G_E$ and $G_C$, 
$r_E^2$ and  $r_C^2$, respectively, are large, revealing 
the long extension of the pion cloud.
We conclude also that those square radii   
are related by $r_E^2 - r_C^2 = 0.6 \pm 0.2$ fm$^2$.
The last result is mainly the consequence 
of the pion cloud effects and Siegert's theorem.
}

%
\maketitle

\section{Introduction}

Among all the nucleon excitations the $\Delta(1232)$
plays a special rule, not only because it is 
a spin 3/2 system with the same quark content as the nucleon,
but also because it is very well known 
experimentally~\cite{NSTAR,Pascalutsa07b,Aznauryan12b}.
The electromagnetic transition between the nucleon ($N$)
and the $\Delta(1232)$
is characterized by the  magnetic
dipole form factor ($G_M$) and two quadrupole form factors:
the electric ($G_E$) and the Coulomb ($G_C$) form factors~\cite{Jones73}.
The magnetic dipole is the dominant form factor,
as expected from the quark spin-flip 
transition~\cite{Beg64,Becchi65,JDiaz07,NDelta,Lattice,OctetDecuplet1,OctetDecuplet2,Eichmann12,Segovia13,SAlepuz17},
while the quadrupole form factors are small, 
but non-zero~\cite{NSTAR,Pascalutsa07b,Isgur82,Capstick90,NDeltaD,LatticeD,Siegert-ND,Letter}.
The transition form factors are traditionally 
represented in terms of $Q^2= -q^2$,
where $q$ is the  $\gamma^\ast N \to \Delta(1232)$ 
momentum transfer, or photon momentum.
Experimental data are available only in the region $Q^2 \ge 0$.

The non-zero results for the quadrupole form factors 
are the consequence of asymmetries on the $\Delta(1232)$
structure, which implies the  deviation of the $\Delta(1232)$
from a spherical shape~\cite{Pascalutsa07b,Becchi65,Capstick90,Glashow79,Bernstein03,Buchmann97a,Krivoruchenko91,Buchmann00b,Deformation,Quadrupole1,Quadrupole2}.
Estimates of the electric and Coulomb quadrupole form factors 
based on valence quark degrees of freedom 
can explain in general only a small fraction 
of the observed data~\cite{Pascalutsa07b,Becchi65,JDiaz07,Capstick90,Buchmann00b,Tiator04,Kamalov99,Kamalov01,SatoLee}.
The strength, missing in quark models can be explained
when we take into account
the quark-antiquark effects in the form of 
meson cloud contributions~\cite{Tiator04,Kamalov99,Kamalov01,SatoLee,Pascalutsa07a,QpionCloud1,QpionCloud2,NSTAR2017}.

Calculations based on non relativistic $SU(6)$ 
quark models with symmetry breaking 
and large $N_c$ limit show that 
the $\gamma^\ast N \to \Delta(1232)$ 
quadrupole form factors are dominated by pion cloud effects at small 
$Q^2$ ($Q^2 < 1$ GeV$^2$)~\cite{Pascalutsa07b,Buchmann97a,Pascalutsa07a,Buchmann04,Grabmayr01,Buchmann09a,Buchmann02}.
Simple parametrizations of the pion cloud contributions to 
the quadrupole form factors $G_E$ and $G_C$,
labeled as $G_E^\pi$ and $G_C^\pi$, respectively,
have been derived using the large $N_c$ limit~\cite{Pascalutsa07a},
in close agreement with the empirical data~\cite{Pascalutsa07a,Blomberg16a}.
Those parametrizations relate $G_E^\pi$ and $G_C^\pi$ 
with the neutron electric form 
factor~\cite{Pascalutsa07a,Grabmayr01,Buchmann09a},
and are discussed in sect.~\ref{secPC}.

There are some limitations associated with 
the use of those pion cloud parametrizations:
they underestimate the $Q^2 <0.2$ GeV$^2$ data
by about 10-20\%~\cite{Siegert-ND,Letter,Blomberg16a,SiegertD},
and they are in conflict with Siegert's theorem, 
a fundamental constraint between the quadrupole form factors 
also known as the 
long wavelength limit~\cite{Siegert-ND,Letter,Buchmann98,Drechsel2007,Tiator1,Tiator2}.  

Siegert's  theorem states that at the 
pseudothreshold, when 
\mbox{$Q^2= Q^2_{pt} \equiv -(M_\Delta-M)^2$,}
one has~\cite{Jones73,Siegert-ND,Letter,SiegertD,Siegert} 
\ba
G_E (Q_{pt}^2)= \kappa G_C(Q^2_{pt}),
\label{eqSiegert1}
\ea
where $M$ and $M_\Delta$ are the nucleon and $\Delta(1232)$ 
masses, respectively, and $\kappa = \sfrac{M_\Delta -M}{2 M_\Delta}$.
The pseudothreshold is the point where the photon three-momentum ${\bf q}$,
vanishes ($|{\bf q}|=0$), and the nucleon and the $\Delta(1232)$ 
are both at rest.

Concerning Siegert's theorem, the problem 
can be sol-ved correcting the parametrization 
for $G_E^\pi$ with a term ${\cal O}(1/N_c^2)$ 
at the pseudothreshold,  as shown recently in~\cite{Letter}.
Concerning the underestimation of the data 
associated with the quadrupole form factors $G_E$ and $G_C$, 
it can be partially solved with 
the addition of contributions 
to those form factors associated with the valence quarks.
Although small, those contributions move 
the estimates based on the pion cloud effects 
in the direction of the measured data~\cite{JDiaz07,LatticeD,Siegert-ND,Kamalov99,Kamalov01}.

In the previous picture, there is only a small setback,
the combination of pion cloud and valence quark 
contributions is unable to describe the previous  
Coulomb quadrupole form factor data below 0.2 GeV$^2$~\cite{Siegert-ND}.
However, it has been shown recently that the previous extractions 
of $G_C$ data at low $Q^2$ ($Q^2 < 0.2$ GeV$^2$) 
overestimate the actual values of the 
form factor~\cite{Blomberg16a}, and that the new results 
are in excellent agreement with the estimates based 
on the pion cloud parametrizations~\cite{Letter}.

The main motivation for this work is to check if there are 
parametrizations of $G_{En}$ which optimizes the description 
of the $G_E$ and $G_C$ data and provides at the same time 
an accurate description of the neutron electric form factor data.
With this goal in mind, we perform a  global fit of the functions 
$G_{En}$, $G_E^\pi$ and $G_C^\pi$, to the 
experimental form factor data,
based on the pion cloud parametrization of the quadrupole form factors.
An accurate  description of the data 
indicates a correlation between  the pion cloud effects 
in the neutron and in the $\gamma^\ast N \to \Delta(1232)$ 
transition~\cite{Pascalutsa07a}.
In addition, a global fit, which includes the $G_E$ and $G_C$ data,
can help to constrain the shape of $G_{En}$,
since the neutron electric form factor data 
have in general large error bars.

The main focus of the present work is then 
on the parametrization of the function $G_{En}$. 
In the past, simplified parametrizations of $G_{En}$ 
with a small number of parameters have been used.
Examples of these parametrizations, are, the Galster parametrization, 
and the Bertozzi parametrization,
based on the differences of two poles,  
among others~\cite{Galster71,Kelly02,Friedrich03,Bertozzi72,Platchkov90,Kaskulov04,Gentile11}.
More sophisticated parametrizations, with a larger number of parameters
have been derived based on dispersion relations and
chiral perturbation theory~\cite{Kaiser03,Belushkin07,Hammer04,Lorenz12}.
For the purpose of the present work, however, 
it is preferable to use simple expressions
with a small number of parameters.
We propose then a parametrization of $G_{En}$
with three parameters.

The parametrizations of $G_{En}$ can also be analyzed in terms of 
first moments of the expansion of $G_{En}$ near \mbox{$Q^2=0$},
namely, the first ($r_n^2$), the second ($r_n^4$) 
and the third ($r_n^6$) moments.
The explicit definitions are presented in sect.~\ref{secGEn}. 
In the present work, motivated by a previous study 
of the $\gamma^\ast N \to \Delta(1232)$ quadrupole form factors 
and their relations with Siegert's theorem~\cite{Siegert-ND},
we consider a parametrization of $G_{En}$ based on rational functions.
Those parametrizations have a simple form for low $Q^2$ 
and are also compatible with the expected falloff at large $Q^2$.
Alternative parametrizations based on 
rational functions can be found in~\cite{Siegert-ND,ComptonScattering}.

As mentioned above, the description of the $G_E$ and $G_C$ data 
can be improved when we include additional 
contributions associated with the valence quark component.
Inspired by previous works,
where the valence quark contributions are estimated 
based on the analysis of the results from lattice QCD~\cite{Lattice,LatticeD,Siegert-ND},  
we include in the present study the valence quark contributions
calculated with the covariant spectator quark model from~\cite{LatticeD}.
The model under discussion is covariant and 
consistent with lattice QCD results~\cite{NDelta,LatticeD,Nucleon}.
Since in lattice QCD simulations with large pion masses 
the meson contributions are suppressed, 
the physics associated with the valence quarks 
can be calibrated more accurately
within the model uncertainties~\cite{Lattice}.
An additional advantage in the use of a quark model framework 
is that the  $G_E$ and $G_C$ form factors are identically zero 
at the pseudothreshold, as a consequence of the 
orthogonality between the nucleon and 
$\Delta(1232)$ states~\cite{LatticeD,Siegert-ND}.
This point is fundamental for the consistence 
of the present model with the constraints of Siegert's theorem.

We conclude at the end, that the combination 
of the $G_{En}$, $G_E$ and $G_C$ data does help to 
improve the accuracy of the parametrization of $G_{En}$
(small errors associated with the parameters).
The best description of the data is obtained 
when the second moment of $G_{En}$, $r_n^4$ is about $-0.38$ fm$^4$.
A consequence of this result is that the square radii 
associated with $G_E$ and $G_C$ ($r_E^2$ and $r_C^2$) are large, 
suggesting a long spacial extension 
of the pion cloud (or quark-antiquark distribution) for both form factors.
We conclude also that, as a consequence of the 
pion cloud parametrizations of the $G_E$ and $G_C$ quadrupole form factors, 
one obtains $r_E^2 - r_C^2 = 0.6 \pm 0.2$ fm$^2$.
The uncertainties of the estimates are also discussed.

This article is organized as follows:
In the next section, we discuss the pion cloud parametrizations 
of the $\gamma^\ast N \to \Delta(1232)$ quadrupole form factors.
In sect.~\ref{secGEn}, we discuss our parametrization 
of the neutron electric form factor.
In sect.~\ref{secResults}, we present the results 
of the global fit to the $G_{En}$, $G_E$ and $G_C$ data,
and discuss the physical consequences of the results.
The outlook and conclusions are presented in sect.~\ref{secConclusions}.

\section{Pion cloud parametrization of $G_E$ and $G_C$}
\label{secPC}

The internal structure 
of the baryons can be described using 
a combination $SU(6)$ quark models with two-body exchange currents
and the large $N_c$ limit~\cite{Buchmann04,Grabmayr01,Buchmann09a}. 
The $SU(6)$ symmetry breaking induces an asymmetric distribution 
of charge in the nucleon and $\Delta(1232)$ systems, which is responsible 
for the non-vanishment of the neutron electric form factor
and for the non-zero results for 
the $\gamma^\ast N \to \Delta(1232)$ quadrupole moments~\cite{Buchmann09a}.
Those results can be derived in the context of constituent quark models 
as the Isgur-Karl model~\cite{Isgur81a,IsgurRefs1,IsgurRefs2}, 
and others~\cite{Pascalutsa07b,Buchmann97a,Grabmayr01}.
More specifically, we can conclude, based on the $SU(6)$ symmetry breaking, 
that the $\gamma^\ast N \to \Delta(1232)$
quadrupole moments are proportional 
to the neutron electric square radius ($r_n^2$).
Based on similar arguments, we can also relate 
the electric quadrupole moment of the $\Delta(1232)$ and 
other baryons, with the neutron electric square 
radius~\cite{Buchmann97a,Krivoruchenko91,Buchmann02,Dillon99,Buchmann02b,Buchmann00a}.
From the large $N_c$ framework, we can conclude 
that when $Q^2=0$, one has $\sfrac{G_E}{G_M}= {\cal O} (\sfrac{1}{N_c^2} )$ 
and $G_E = \sfrac{M_\Delta^2 - M^2}{4 M_\Delta^2} G_C$~\cite{Pascalutsa07b,Jenkins02}.

The derivation of the pion cloud contribution 
to the  $\gamma^\ast N \to \Delta(1232)$ quadrupole form factors 
are based on relations between the quadrupole moments
and the neutron electric square radius, discussed above, 
and on the low-$Q^2$ expansion  of the neutron electric form factor:
$G_{En} \simeq -\sfrac{1}{6} r_n^2 Q^2$~\cite{Buchmann97a,Pascalutsa07a,Buchmann04,Grabmayr01,Buchmann09a,Buchmann02}.
One can then write
\ba
\hspace{-.5cm}
& &
G_E^\pi (Q^2) = \left(\frac{M}{M_\Delta} \right)^{3/2} 
\frac{M_\Delta^2 -M^2}{2 \sqrt{2}} 
\frac{\tilde G_{En}(Q^2)}{ 1 + \frac{Q^2}{2 M_\Delta (M_\Delta-M)}}, 
\label{eqGE1} \\
\hspace{-.5cm}
& &
G_C^\pi (Q^2) = \left(\frac{M}{M_\Delta} \right)^{1/2} 
\sqrt{2} M_\Delta M \tilde G_{En}(Q^2), 
\label{eqGC1}
\ea
where $\tilde G_{En} = G_{En}/Q^2$.

The interpretation of the previous relations 
as representative of the pion cloud contributions 
is a consequence of the connection between 
quadrupole form factors and $r_n^2$.
Those relations have been derived in the framework 
of the constituent form factors with two-body exchange currents.
The effects of those currents can be interpreted 
as pion/meson contributions since they include processes
associated with pion exchange and 
quark-antiquark pairs~\cite{Buchmann04,Grabmayr01,Buchmann09a,Buchmann02}.
This interpretation is also valid in the large $N_c$ limit,
where the form factor $G_E$ and the 
product $\frac{|{\bf q}|}{2  M_\Delta} G_C$ 
appear as higher orders in $1/N_c$
compared to $G_M$~\cite{Pascalutsa07a,Jenkins02}.

For future discussions, it is worth noticing that 
the $SU(6)$/large$N_c$ estimates are 
based on non relativistic kinematics and therefore 
may ignore some high angular momentum state effects 
that emerge in a relativistic framework.

Indications of the dominance of the pion cloud 
contributions have been found also within 
the framework of the dynamical coupled-channel models,
like the Sato-Lee and the DMT models~\cite{JDiaz07,Kamalov99,Kamalov01,SatoLee,Blomberg16a}.

It is worth mentioning that eqs.~(\ref{eqGE1})-(\ref{eqGC1}) 
should not be strictly interpreted as pion cloud contributions, 
because the empirical parametrization of $G_{En}$ 
include all possible contributions, including 
also the effects of the valence quarks.
We note, however, that in an exact $SU(6)$ model 
the contributions from the valence quarks 
associated  with one-body currents 
vanishes~\cite{Buchmann09a,Dillon99,Lichtenberg78,Close89}.
In a model where the $SU(6)$ symmetry is 
broken in first approximation, 
one can expect 
that the quark-antiquark contributions are 
the dominant effect in $G_{En}$ and $r_n^2$~\cite{Grabmayr01,Buchmann09a}.
Examples of models with meson cloud/sea quark  dominance can be found 
in~\cite{Buchmann00b,Buchmann97a,Buchmann91,Christov96,Lu98}. 
We conclude then that (\ref{eqGE1})-(\ref{eqGC1})
can still be used to estimate the pion cloud contribution 
to the $\gamma^\ast N \to \Delta(1232)$ 
quadrupole form factors in the cases where 
the valence quark contributions are small.

We assume that eqs.~(\ref{eqGE1})-(\ref{eqGC1})
and in particular the  analytic parametrizations of $G_{En}$ hold also for $Q^2 < 0$.
This assumption is justified by the smooth behavior
inferred for $G_{En}$ based on the empirical data,
and also because the range of extrapolation 
is small, since $Q_{pt}^2 \simeq -0.086$ GeV$^2$.
In this range, the effects associated with the timelike 
pole structure, such as the effects of the $\rho$-pole 
($Q^2 \simeq - 0.6$ GeV$^2$), are attenuated, 
or can be represented in an effective form.
In sect.~\ref{secGEn2} we show that $G_{En}$ is in fact dominated, 
near the pseudothreshold, by the first two terms of the expansion in $Q^2$.
Note that the two pion production threshold 
$Q^2 = -4 m_\pi^2 \simeq -0.076$ GeV$^2$, 
the point where the transition form factors become complex functions, 
is just a bit above the 
$\gamma^\ast N \to \Delta(1232)$ pseudothreshold.
Since, according to chiral perturbation theory, the transition 
to the complex form factors is smooth \cite{Kaiser03,Belushkin07},
we can ignore the imaginary component 
of the form factors in first approximation, 
and treat the form factors in 
the timelike region as simple extrapolations 
of the analytic parametrizations derived in the spacelike region.
Similar extrapolations to the timelike region can also be found 
in~\cite{Drechsel2007,Tiator07,Tiator11}.

The relations (\ref{eqGE1})-(\ref{eqGC1}) are derived in~\cite{Letter}
and improve previous large $N_c$ relations~\cite{Pascalutsa07a},
in order to satisfy Siegert's theorem exactly~(\ref{eqSiegert1}).
The main difference between our analytic 
expressions for the functions $G_E^\pi$ and $G_C^\pi$
compared to the expressions derived in~\cite{Pascalutsa07a}, 
is in eq.~(\ref{eqGE1}), where we include  
a denominator in the factor $\tilde G_{En}$.
This new factor corresponds to a relative correction ${\cal O}(1/N_c^2)$,
relative to the original derivation for $G_E^\pi$, 
at the pseudothreshold\footnote{Since in the large $N_c$ limit 
$M_\Delta -M = {\cal O}(1/N_c)$, and $M$, $M_\Delta = {\cal O}(N_c)$, 
the factor $1/\left(1 + \sfrac{Q^2}{2 M_\Delta(M_\Delta - M)}\right)$ 
corresponds to a correction ${\cal O}(1/N_c^2)$,
at the pseudothreshold.}.
Using the new form, one obtains at the pseudothreshold:
$1 + \sfrac{Q^2}{2 M_\Delta (M_\Delta-M)} = \sfrac{M_\Delta + M}{2M_\Delta}$, 
which lead directly to eq.~(\ref{eqSiegert1}).
In a previous work~\cite{Siegert-ND}, an approximated expression 
was considered for $G_E^\pi$, where the error in 
the description of Siegert's theorem is a term ${\cal O} (1/N_c^4)$.

Previous studies of Siegert's theorem 
based on quark models~\cite{Capstick90,Buchmann98,Drechsel84,Weyrauch86,Bourdeau87} 
show that the theorem 
can be violated when the operators associated 
with the charge density, or the current density, are truncated 
in different orders, inducing a violation of the 
current conservation condition~\cite{Buchmann98}.
From those works, we can conclude that, 
a consistent calculation with current conservation,
cannot be reduced to the photon coupling 
with the individual quarks (one-body currents), 
since the current is truncated.
In those conditions, it is necessary the inclusion of 
higher-order terms such as two-body currents,
in order to ensure current conservation 
and to be consistent with Siegert's theorem~\cite{Buchmann98}.
The description of Siegert's theorem,
requires then the inclusion of processes beyond the impulse approximation 
(one-body currents)~\cite{SiegertD,Buchmann98}.

We recall at this point that eqs.~(\ref{eqGE1})-(\ref{eqGC1}) 
are the result of derivations valid at small $Q^2$~\cite{Pascalutsa07a}.
For very large values of $Q^2$, we expect  
the pion cloud contributions 
to be negligible in comparison 
to the valence quark contributions.
Thus, one can assume that for large $Q^2$,
the equations (\ref{eqGE1})-(\ref{eqGC1}) are corrected 
according to $G_E^\pi \to G_E^\pi/(1+ Q^2/\Lambda_E^2)^2$
and $G_C^\pi \to G_C^\pi/(1+ Q^2/\Lambda_C^2)^4$, 
where $\Lambda_E$ and $\Lambda_C$ 
are large momentum cutoff parameters~\cite{Letter}.
In those conditions, the form factors $G_E$ 
and $G_C$ are, at large $Q^2$, dominated by the 
valence quark contributions, as predicted 
by perturbative QCD, with  falloffs 
$G_E \propto 1/Q^4$ and $G_C \propto 1/Q^6$~\cite{Carlson1,Carlson2}.
The falloffs of the pion cloud components 
are then $G_E^\pi \propto 1/Q^8$ and $G_C^\pi \propto 1/Q^{10}$,
where the extra factor ($1/Q^4$) takes into account 
the effect of the extra quark-antiquark pair
associated with the pion cloud~\cite{Carlson1,Carlson2}.

\begin{figure}[t]
\vspace{.7cm}
\centerline{\mbox{
\includegraphics[width=3.0in]{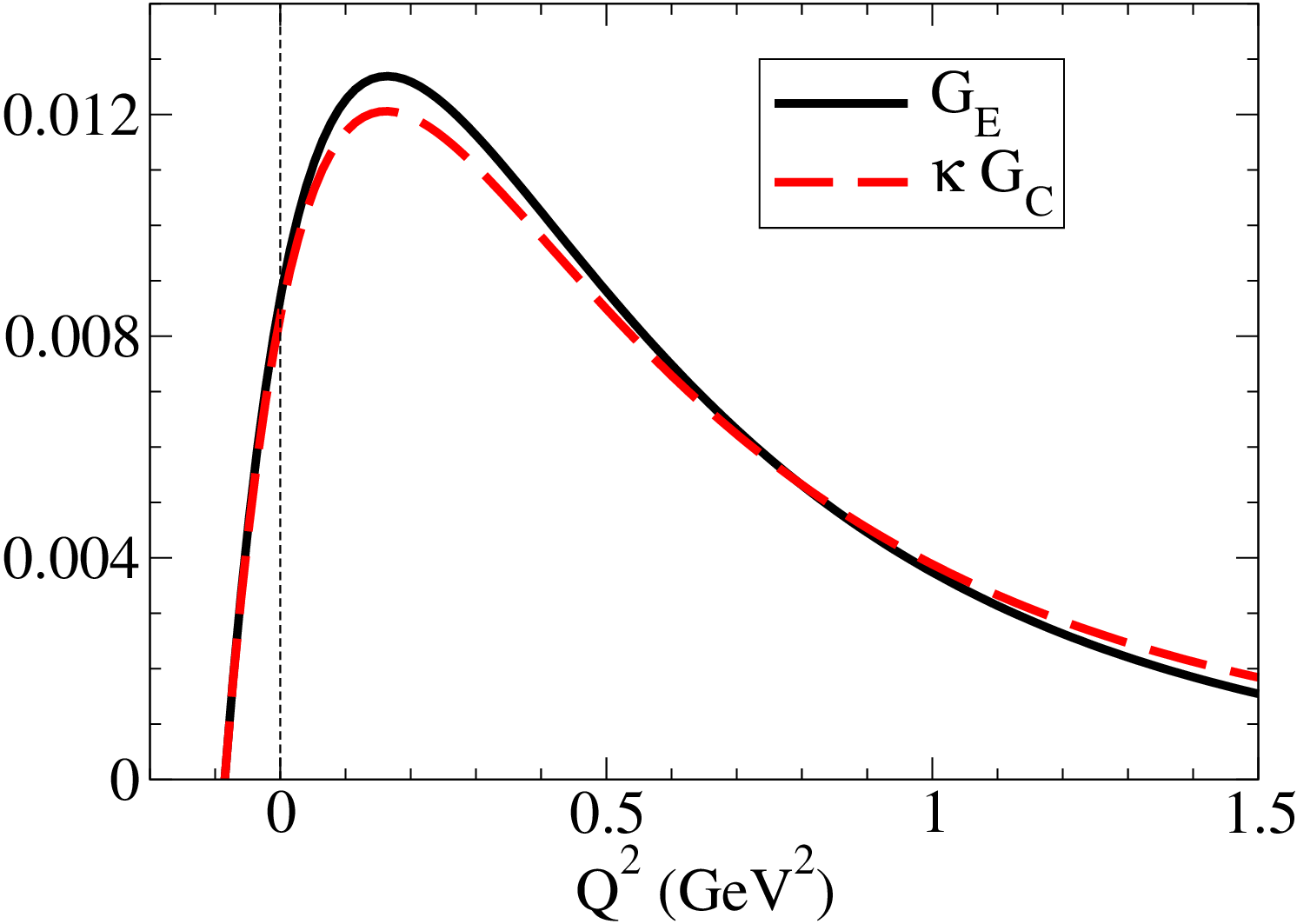}}}
\caption{\footnotesize
Valence quark contributions for $G_E$ and $G_C$ 
according to~\cite{LatticeD}.
Recall that $\kappa = \sfrac{M_\Delta-M}{2M_\Delta} \simeq 0.12$.}
\label{figValence}
\end{figure}

\subsection*{Adding the valence quark contributions}

The pion cloud contributions to
the  $\gamma^\ast N \to \Delta(1232)$ quadrupole form factors 
given by eqs.~(\ref{eqGE1})-(\ref{eqGC1}) 
should be complemented by  
valence quark contributions to the respective form factors.
Calculations from quark models
estimate that the valence quark contribution 
to the $\gamma^\ast N \to \Delta(1232)$
quadrupole form factors are an order of magnitude 
smaller than the 
data~\cite{NSTAR,JDiaz07,Capstick90,NDeltaD,NSTAR2017,Stave08}.
It is known, however, that small valence quark contributions to the quadrupole 
form factors can help to improve 
the description of the data~\cite{LatticeD,Siegert-ND}.

The valence quark contributions 
to the $\gamma^\ast N \to \Delta(1232)$
quadrupole form factors are in general the result 
of high angular momentum components 
in the nucleon and/or $\Delta(1232)$ wave functions.
The valence quark contributions to the quadrupole 
form factors vanish at the pseudothreshold,
as a consequence of the orthogonality between 
the nucleon and  $\Delta(1232)$ states,
as discussed in~\cite{Siegert-ND}.
In those conditions, the valence quark contributions 
to the quadru-pole form factors play no role in  Siegert's theorem,
which depends exclusively on the pion cloud contributions.
The parameterizations (\ref{eqGE1})-(\ref{eqGC1}) then 
ensure that Siegert's theorem 
is naturally satisfied.

In order to improve the description 
of the $\gamma^\ast N \to \Delta(1232)$ quadrupole form factors,
we combine the pion cloud  parametrizations (\ref{eqGE1})-(\ref{eqGC1})
with the valence quark contributions (index B), 
using $G_a= G_a^{\rm B} + G_a^\pi$, for  $a=E,C$. 
For the valence quark contributions 
we use in particular the quark model from~\cite{LatticeD},
since in this work the valence quark contributions 
are calculated using an extrapolation from 
the lattice QCD simulations 
in the quenched approximation
with large pion masses~\cite{Alexandrou08}.
Recall that in lattice QCD simulations with large pion masses, the
meson cloud effects are very small, and the physics associated
with the valence quarks can be better calibrated~\cite{Lattice,LatticeD,Omega}.
The free parameters of the model are 
two $D$-state admixture coefficients and three momentum  
scales associated with the radial structure of the $D$-states.
All those parameters are determined accurately 
by the results from lattice QCD simulations~\cite{LatticeD}.
The chi-square per degree of freedom   
associated with the fit of the model 
to the lattice QCD results is 0.58 for 
$G_E$ and $1.17$ for $G_C$.
The number of degrees of freedom of the fit is 
37 and the combined chi-square per degree of freedom is 0.76.

The model under discussion is covariant and can therefore 
be used to estimate the valence quark contributions 
in any range of $Q^2$~\cite{LatticeD,Nucleon}.
The results of the valence quark contributions 
to the form factors $G_E$ and $G_C$ estimated by the model 
from~\cite{LatticeD} are presented in fig.~\ref{figValence}.
The available data near $Q^2=0$ 
is $G_E(0) = 0.076 \pm 0.015$ and 
$\kappa G_C(0.04\; \mbox{GeV}^2) = 0.075 \pm 0.021$~\cite{Letter,Blomberg16a,PDG}.
Based on fig.~\ref{figValence},
one can then conclude that the valence quarks contribute  
only for about 10\% of the empirical data near $Q^2=0$.
The theoretical uncertainties associated 
with the results from  fig.~\ref{figValence}
are discussed later on.

Estimates based on the Dyson-Schwinger formalism 
suggest that relativistic effects, 
in particular $P$-state contributions, 
can increase the magnitude 
of the valence quark contributions~\cite{Eichmann12,Segovia13}.
These results suggest that our 
estimates of the pion cloud contributions may be 
overestimated.

\section{Parametrizations of $G_{En}$}
\label{secGEn}

The parametrizations of $G_{En}$ 
can be represented by an expansion near $Q^2=0$ in the form~\cite{Grabmayr01}:
\ba
G_{En} (Q^2) \simeq - \sfrac{1}{6} r_n^2 Q^2 +  \sfrac{1}{120} r_n^4 Q^4 
-  \sfrac{1}{5040} r_n^6 Q^6 +
...,
\label{eqGEn-expan}
\ea
where each term defines a momentum of the function  $G_{En}$.
In the previous equation, $r_n^2$ is the first moment, 
$r_n^4$ is the second moment, $r_n^6$ is the third moment, and so on.
The first moment is well known experimentally:
$r_n^2 \simeq - 0.116$ fm$^2$~\cite{PDG}.
The value of the second moment ($r_n^4$) controls 
the curvature of the function $G_{En}$ near $Q^2=0$.
For $r_n^4$, we need, at the moment, to rely on models.
A discussion about the possible values for
$r_n^4$ can be found in~\cite{Grabmayr01}.
The third moment ($r_n^6$) is usually omitted in the discussions,
it can be used to discriminate two close parametrizations.

For the discussion, it is important to 
present also the form of the most popular parametrization of $G_{En}$,
the Galster parametrization~\cite{Galster71,Kelly02,Platchkov90}:
\ba
G_{En}(Q^2) = - \frac{1}{6}r_n^2 \frac{Q^2}{1 + d\, \tau_{\scriptscriptstyle N}} G_D,
\label{eqGalster}
\ea
where  $\tau_{\scriptscriptstyle N}= \frac{Q^2}{4M^2}$, 
$d$ is a free parameter, 
and $G_D$ is the nucleon dipole 
\mbox{$G_D=1 /(1 + Q^2/\Lambda^2 )^{2}$}
with $\Lambda^2 =0.71$ GeV$^2$.
Since $r_n^2$ is well determined experimentally,
there are only two independent parameters in (\ref{eqGalster}),
counting $\Lambda$ as a parameter.

In the present work, we use the following 
form to para-metrize the function $\tilde G_{En}$:
\ba
\tilde G_{En} (Q^2)= 
\frac{c_0(1  + c_2 Q^2 + ...  + c_k Q^{2k-2})}{1+ c_1 Q^2 + \sfrac{c_1^2}{2!}Q^4 
+ ... +  \sfrac{c_1^{k+2}}{(k + 2)!}Q^{2k +4}  },
\label{eqGEntil1}
\ea
where $k=2,3,...$ is an integer and 
$c_l$ ($l=0,...,k$) are adjustable coefficients.
The use of eq.~(\ref{eqGEntil1}) is motivated 
by the relation between the quadrupole form factors 
$G_E$ and $G_C$ and the neutron electric form factor $G_{En}$,
and also by previous studies of Siegert's theorem~\cite{SiegertD}.
The parameterizations of the transition form factors are sometimes
performed based on analytic forms that are valid only 
in a limited region of $Q^2$. 
Examples of those representations are the MAID parametrizations,
based on polynomials and exponentials~\cite{Drechsel2007}.
Those representations have problems in the extension 
for the timelike region (large exponential effects)
and in the large-$Q^2$ region where the form factors 
have very fast exponential falloffs, instead of  
power law falloffs predicted by perturbative QCD~\cite{Carlson1,Carlson2}.
Alternatively, the parametrizations based on rational functions 
can describe both the low- and large-$Q^2$ regions with the 
appropriate power law falloffs, 
and can also be extended to the timelike region.
Depending on the resonance under study, 
the parameter $c_1$ can be chosen in order to avoid 
singularities above the pseudothreshold~\cite{SiegertD}.
Note that for large $Q^2$, one has $\tilde G_{En} \propto 1/Q^6$,
as expected from perturbative QCD arguments~\cite{Carlson1,Carlson2}.

The form (\ref{eqGEntil1}) was used in~\cite{SiegertD}
to parametrize directly the $G_C$ data.
A modified form with a different asymptotic falloff was also
used to parametrize the $G_E$ data ($G_E \propto 1/Q^4$, for large $Q^2$).
From eq.~(\ref{eqGEntil1}), one can derive 
the following relations for  
the first ($r_n^2$), the second ($r_n^4$), 
and the third ($r_n^6$) moments of $G_{En}$, 
as defined in eq.~(\ref{eqGEn-expan}):
\ba
& &
c_0 = - \frac{1}{6} r_n^2, 
\nonumber \\
& &
c_0(c_2-c_1) =  \frac{1}{120} r_n^4,  \nonumber \\ 
& &
\frac{1}{2}c_0\left[c_1(c_1-c_2)  + 2c_3\right] =  - \frac{1}{5040} r_n^6.
\ea
We fix the value of $r_n^2$ by the experimental value 
$r_n^2 \simeq -0.116$ fm$^2$ (or $-2.98$ GeV$^{-2}$),
due to its precision.
Typical values for $r_n^4$ are 
in the range 
$r_n^4 = -$(0.60--0.30) fm$^4$~\cite{Grabmayr01,Platchkov90,Kaskulov04}.
The third moment is discussed later. 

The parametrization (\ref{eqGEntil1}) 
has only $k$ independent coefficients: $c_l$ ($l=1,...,k$),
because $c_0$ is fixed by the data.
There are then $k$ adjustable parameters.


\section{Combined fit of $G_{En}$, $G_E$ and $G_C$}
\label{secResults}

In this section, we present the results of the global 
fit of eqs.~(\ref{eqGE1}), (\ref{eqGC1}) and (\ref{eqGEntil1})
to the $G_{En}$, $G_E$ and $G_C$ data, based on the chi-square minimization.  
Since the pion cloud expressions for $G_E$ and $G_C$ 
are derived at low $Q^2$~\cite{Pascalutsa07a}, 
we restrict the fit of the quadrupole form factors 
to the $Q^2 < 1.5$ GeV$^2$ region.
For $G_{En}$ we consider no restrictions in the data.
The quality of the fits is measured by the 
chi-square value per degree of freedom 
(reduced chi-square) associated with 
the different data sets.

We start by discussing the data used in our fits
and the explicit form used for $G_{En}$ 
(sects.~\ref{secData} and \ref{secGEn1}).
After that, we present the results for two different cases.
First, we consider a simple model 
where we neglect 
the effect of the valence quark contributions (sect.~\ref{sec-MinimalModel}).
Next, we present the results of our best fit to the global data,
including the valence quark contributions 
to the $\gamma^\ast N \to \Delta(1232)$ quadrupole form factors 
(sect.~\ref{sec-GlobalFit}).
Once determined our best fit to the global data 
we discuss the results for $G_{En}$ (sect.~\ref{secGEn2})
and the results to the 
$\gamma^\ast N \to \Delta(1232)$ quadrupole form factors
(sect.~\ref{secQ-Delta}),
including the values obtained for the 
electric and Coulomb quadrupole square radii (sect.~\ref{secRadii}).
At the end, we debate some theoretical and 
experimental aspects related to the final results.

\subsection{Data}
\label{secData}

In the present work, we consider the data for $G_{En}$ 
from~\cite{Gentile11,Schiavilla01}.
The data from~\cite{Schiavilla01} are extracted from 
the analysis of the deuteron quadrupole form factor data,
providing data at very low $Q^2$.
Reference~\cite{Gentile11} presents a compilation 
of the data from different double-polarization 
experiments~\cite{Eden94,Passchier99,MainzR1,MainzR2,Bermuth03,JlabR2,JlabR3,JlabR4,Geis08,Riordan10}, 
which measure the ratio $G_{En}/G_{Mn}$,
where $G_{Mn}$ is the neutron magnetic form factor.
In total, we consider 35 data points for $G_{En}$.

Concerning the $\gamma^\ast N \to \Delta(1232)$ quadrupole form factor data, 
we consider a combination of the database from \cite{MokeevDatabase},
the recent data from JLab/Hall A~\cite{Blomberg16a},
and the world average of the Particle Data Group for $Q^2=0$~\cite{PDG}. 
The database from~\cite{MokeevDatabase}
includes data from MAMI~\cite{Stave08},
MIT-Bates~\cite{MIT_data} and JLab~\cite{Jlab_data1,Jlab_data2} 
for finite $Q^2$.
Comparatively to~\cite{MokeevDatabase},
we replace the $G_C$ data below 0.15 GeV$^2$
from~\cite{Stave08,MIT_data}, by the new data 
from~\cite{Blomberg16a}.
This procedure was adopted because it was shown 
that there is a discrepancy between the new data and previous measurements
from MAMI and MIT-Bates~\cite{Stave08,MIT_data} below 0.15 GeV$^2$,
due to differences in the extraction procedure of the resonance 
amplitudes from the measured cross sections~\cite{Blomberg16a}.
For the same reason, the MAMI data from~\cite{Sparveris13} 
are not included. 
As for $G_E$, we combine the data from~\cite{MokeevDatabase}
with the more recent data from JLab/Hall A~\cite{Blomberg16a}.
Since we restrict the quadrupole form factor data 
to the region $Q^2 < 1.5$ GeV$^2$,
the $G_E$ data is restricted to 15 data points 
and the $G_C$ data to 13 data points.

The data for $G_E$ and $G_C$ are converted 
from the results from~\cite{Blomberg16a,MokeevDatabase,PDG}.
In the case of~\cite{MokeevDatabase} 
the form factors $G_M$, $G_E$ and $G_C$ are calculated 
from the helicity amplitudes $A_{1/2}$, $A_{3/2}$ and $S_{1/2}$ 
using standard relations~\cite{Pascalutsa07b,Aznauryan12b,Drechsel2007}.
For $Q^2=0$, we use the PDG data~\cite{PDG} for  
the electromagnetic ratios $R_{EM} \equiv - \frac{G_E}{G_M}$
and $R_{SM} \equiv - \frac{|{\bf q}|}{2 M_\Delta}  \frac{G_C}{G_M}$,
combined with the experimental value of $G_M(0)$~\cite{Drechsel2007}
(${\bf q}$ is the photon three-momentum 
in the $\Delta(1232)$ rest frame).
As for the recent data from~\cite{Blomberg16a} 
for $R_{EM}$ and $R_{SM}$, we use the MAID2007 parametrization for $G_M$, 
since the helicity amplitudes are not available for all values of $Q^2$.
The MAID2007 parametrization for $G_M$ can be expressed as 
\mbox{$G_M = 3 \sqrt{1 + \tau} ( 1 + a_1 Q^2) e^{-a_4 Q^2}$,}
where $\tau= \frac{Q^2}{(M_\Delta + M)^2}$, 
$a_1 =0.01$ GeV$^{-2}$ and $a_4 = 0.23$ GeV$^{-2}$~\cite{Drechsel2007}.
The MAID-2007  parametrization 
provides a very good description of the low-$Q^2$ data.

\begin{figure}[t]
\vspace{.6cm}
\centerline{\mbox{
\includegraphics[width=3.0in]{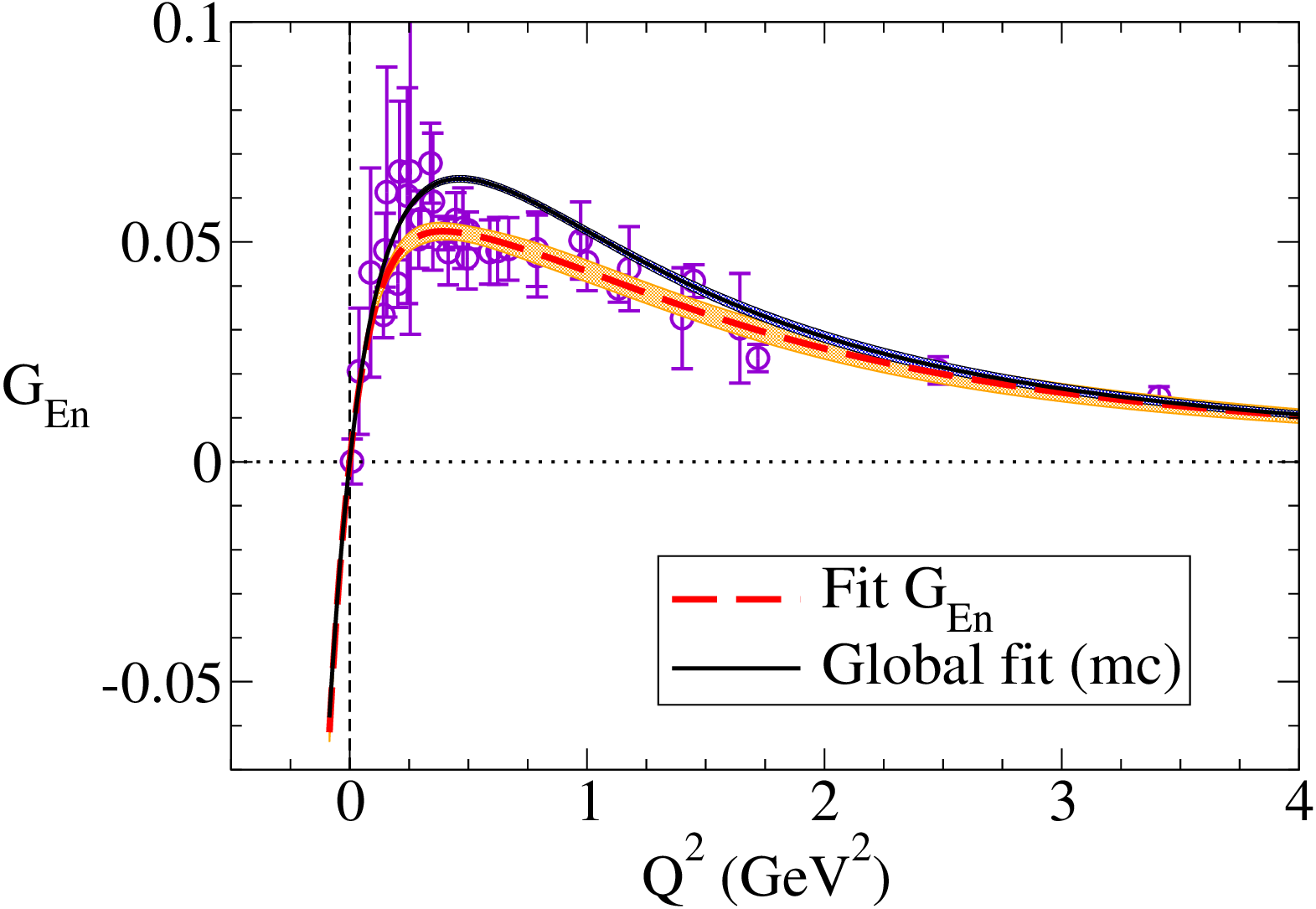}}}
\caption{\footnotesize
Neutron electric form factor. 
Results of the direct fit to the $G_{En}$ data (dashed line, thick band),
and of the global fit without the valence quark contributions (thin band).
The label ``mc'' stands for meson cloud.
Data from~\cite{Gentile11,Schiavilla01}.}
\label{figGEn-v1}
\end{figure}

\begin{figure*}[t]
\vspace{.6cm}
\centerline{
\mbox{\includegraphics[width=3.in]{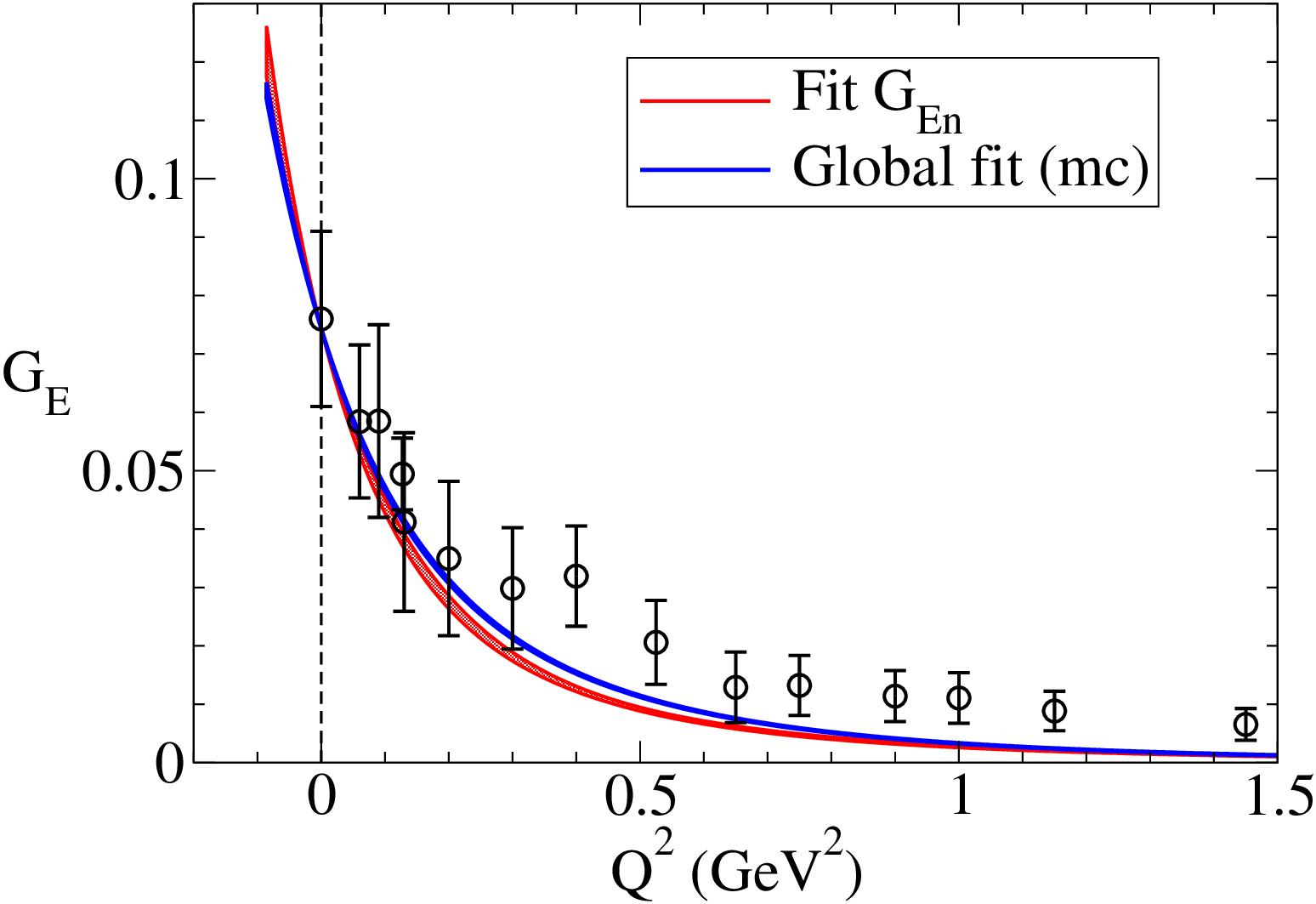}} \hspace{.8cm}
\mbox{\includegraphics[width=3.in]{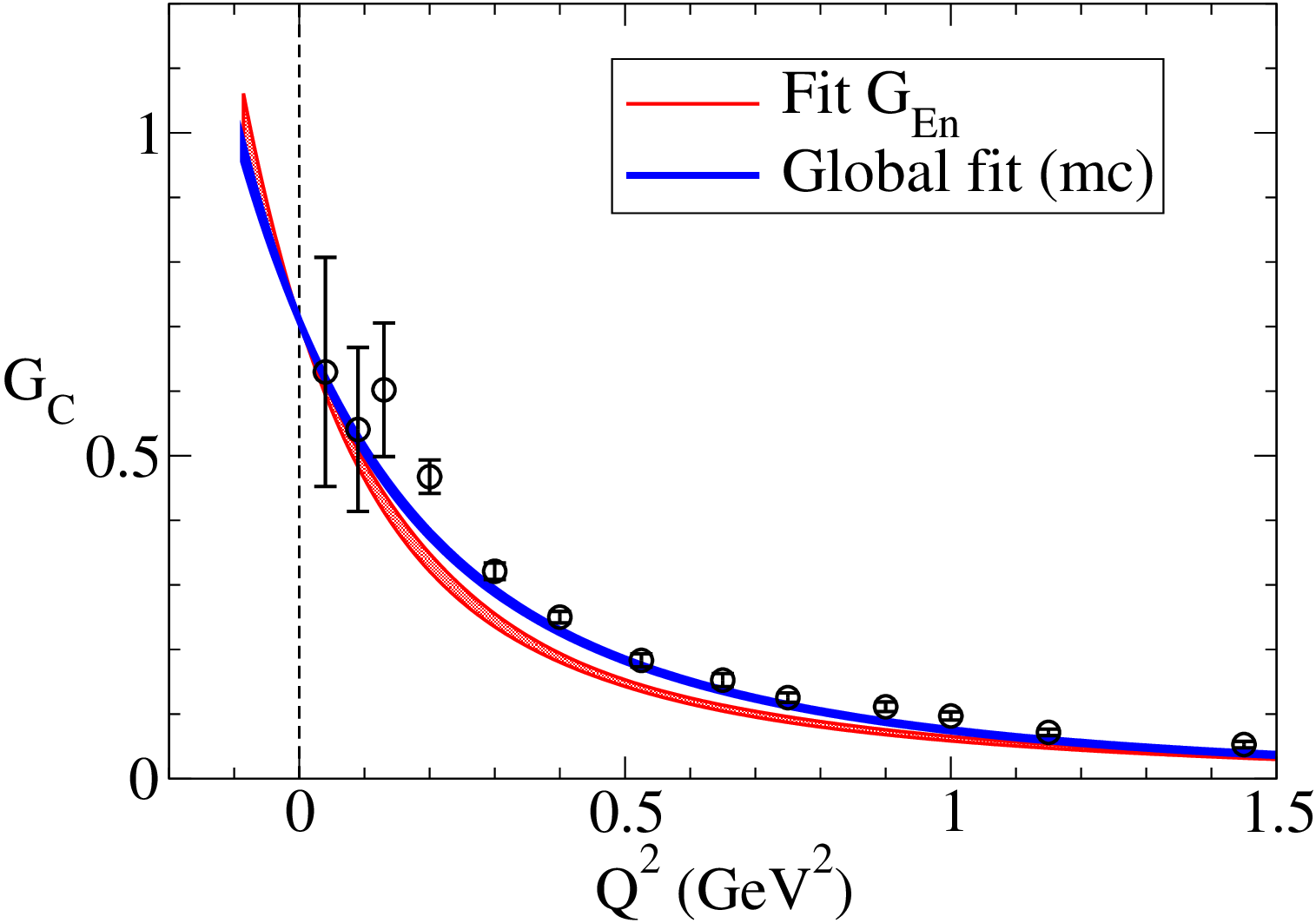}}}
\caption{\footnotesize
$\gamma^\ast N \to \Delta(1232)$ electric and
Coulomb quadrupole form factors. 
Results of models with no valence quark contributions.
Results of the direct fit to the $G_{En}$ data (red band),
and the global fit (blue band).
The label ``mc'' stands for meson cloud.
Data for proton targets 
from~\cite{Blomberg16a,Stave08,PDG,MIT_data,Jlab_data1,Jlab_data2}.}
\label{figGEGC-v1}
\end{figure*}

\subsection{Parametrization for $G_{En}$}
\label{secGEn1}

In the calculation of the pion cloud contributions  
to $G_E$ and $G_C$, we use the parametrization 
of $G_{En}$ given by eq.~(\ref{eqGEntil1}), 
where we take $k=3$ (3 parameters).
Compared to the Galster parametrization (\ref{eqGalster}),
we consider one additional parameter.
Using eq.~(\ref{eqGEntil1}), 
we avoid an explicit dependence on the dipole factor $G_D$.
This procedure may be more appropriated for 
the study of the $\gamma^\ast N \to \Delta(1232)$ 
transition form factors, since it avoids 
the connection with a scale associated 
exclusively with the 
nucleon~\cite{Siegert-ND}.

Fits based on parametrizations with larger values of $k$ 
($k \ge 4$) lead to similar 
results at low $Q^2$, but increase the number of parameters.
In addition, for $k \ge 5$, the higher order coefficients $c_l$ 
of the parametrizations are constrained mainly by the  $Q^2 > 2$ GeV$^2$ data, 
which is restricted to 2 data points (2.48 GeV$^2$ and  3.41 GeV$^2$).
Those coefficients are therefore poorly constrained,
and can lead to strong oscillations in the function $G_{En}$ for large $Q^2$, 
in conflict with the smooth falloff expected for large $Q^2$.
Parametrizations of $G_{En}$ with $k \ge 5$  may be considered, 
once more information relative the function $G_{En}$ become available,
such as, more and better distributed data at large $Q^2$.


\begin{table*}[t]
\caption{
Parameters $c_i$ associated with the parametrization 
from eq.~(\ref{eqGEntil1}) for the case $k=3$.
The brackets represent the uncertainty associated with 
the parameters.
In the first row $\rho_{c_i c_j}$ represents 
the coefficient of correlation between $c_i$ and  $c_j$.
In the second row $\sigma_{c_i c_j}$ 
represent the covariance between $c_i$ and $c_j$.}
\label{tabMod0}
\begin{center}
\begin{tabular}{l| c c c c c c}
\hline
\hline
  & $c_1$ (GeV$^{-2}$)&  $c_2$ (GeV$^{-2}$) &  $c_3$ (GeV$^{-4}$) 
  &   $\rho_{c_1 c_2}$  &   $\rho_{c_1 c_3}$   &
  $\rho_{c_2 c_3}$ \\
  &                  &                    & 
  & $\sigma_{c_1 c_2}$ (GeV$^{-4}$) &  $\sigma_{c_1 c_3}$ (GeV$^{-6}$)  & $ \sigma_{c_2 c_3}$ (GeV$^{-6}$) \\
\hline
Fit $G_{En}$ & 3.39   &  $-0.750$  &    2.01  & 0.39  &  0.97  & 0.21\\  
              & (0.39) &  (0.135) &   (0.77)  & (0.021) & (0.29) & (0.025)\\
\hline
Global fit          &   3.31   &  $-0.175$  &  1.71   &  0.84   &  0.97  & 0.69 \\  
(only meson cloud)  &  (0.19)  &  (0.113)   & (0.34)  & (0.019) &(0.063) & (0.026) \\
\hline
{\bf Global fit} &            3.28   &  $-0.954$  &   1.93  & 0.62  &  0.95  & 0.36 \\  
{\bf (total)}             &  (0.17)  &  (0.077) &  (0.26) &(0.0079) &(0.042) & (0.0073) \\
\hline
Fit $G_{En}$, $G_E$        &   3.47 & $-0.744$ & 2.19 &   0.35  &  0.98 & 0.19 \\
                          &  (0.42) & (0.138) & (0.90) & (0.020) & (0.37) & (0.015) \\ 
\hline
\hline
\end{tabular}
\end{center}
\end{table*}

\subsection{Model with no valence quark contributions}
\label{sec-MinimalModel}

In order to test  whether 
the inclusion of the valence quark effects on the 
$\gamma^\ast N \to \Delta(1232)$ quadrupole form factors 
is really relevant for the description of the data,
we consider first two fits where we ignore those effects.

We start with a direct fit to the $G_{En}$ data, 
ignoring the impact of the quadrupole form factor data.
The parameters associated with the fit 
and the respective errors 
are presented in the first row of table~\ref{tabMod0}.
The coefficients of correlation between 
the parameters are also 
presented in the table and are discussed later.
The reduced chi-square associated with the $G_{En}$ fit is 0.66, 
pronouncing the good quality of fit.
The parameters of the global fit to the $G_{En}$, $G_E$ and $C_C$ data,
using only the meson cloud term, 
and the corresponding errors are presented in 
the second row of table~\ref{tabMod0}.
The values of reduced chi-square 
associated with the different subsets and 
the combined chi-square per degree of freedom 
are presented in the second row of table~\ref{tabChi2a}.

From the chi-square values presented in
the second row of table~\ref{tabChi2a}, one can anticipate 
that a fit that does not include additional contributions 
to the pion cloud 
parametri-zation lead to a very poor description of 
the $G_C$ data (reduced chi-square 6.7), 
as well as an imprecise description of the 
neutron electric form factor data (reduced chi-square 2.7)
and the electric quadrupole form factor data (reduced chi-square 2.0).

The results of the two fits to the $G_{En}$, 
including the interval 
of variation associated with the  
uncertainties in the parameters are presented
in fig.~\ref{figGEn-v1}.
The dashed line represent the result 
of the fit to the $G_{En}$ data
and the uncertainties are 
represented by the thick band (orange).
The thin band represents the result of the global fit
(including the uncertainties).
The range of the variation of $G_{En}$ 
is estimated using the propagation
of errors associated with the coefficients $c_i$ 
according to the parametrization (\ref{eqGEntil1}) with $k=3$~\cite{Errors-Taylor}.

In fig.~\ref{figGEn-v1}, we also present the results 
below $Q^2=0$ down to $Q^2= - (M_\Delta -M)^2 \simeq -0.09$ GeV$^2$,
which correspond to the pseudothreshold of the 
$\gamma^\ast N \to \Delta(1232)$ transition.
As mentioned, the region below  $Q^2=0$ is important 
for the study of the quadrupole form factors 
and to Siegert's theorem~\cite{Siegert-ND,Letter}.

The results for the quadrupole form factors, $G_E$ and $G_C$,
are presented in fig.~\ref{figGEGC-v1}.
In this case, the interval of variation is small (narrower bands).

The errors associated with the coefficients $c_j$ are 
determined based on standard relations that 
take into account the errors associated with 
the data and the sensibility of the parameters 
relative to  the data.
The last effect is  estimated by $\left(\frac{\partial c_j}{\partial y_i}\right)$ 
where $y_i$ represent the 
$G_{En}$, $G_E$ or $G_C$ data~\cite{Errors-Taylor,NumericalRecipes}.
The available data suggest that the 
coefficients $c_1$, $c_2$ and $c_3$ are not really uncorrelated,
since the coefficients of 
correlation between the parameters 
differ significantly from zero~\cite{Errors-Taylor,NumericalRecipes}.
The coefficients of correlation and the covariance functions
are presented in the last three columns of table~\ref{tabMod0}.
Note in particular that the coefficient of correlation 
between $c_1$ and $c_3$, $\rho_{c_1 c_3 }$,
is very close to one, indicating a strong correlation between
the effect of those two coefficients.
%
A consequence of the correlation between coefficients 
is that  to estimate the range of variation of $G_{En}$
we cannot consider just the quadratic terms 
$ \left(\frac{\partial G_{En}}{\partial c_i} \right)^2  \sigma_{c_i}^2$
associated with $c_1$, $c_2$ and $c_3$, but that 
it is also necessary to consider the cross terms: 
$2 \left(\frac{\partial G_{En}}{\partial c_i} \right)  
\left(\frac{\partial G_{En}}{\partial c_j} \right) 
\sigma_{c_i c_j}$, 
where 
$\sigma_{c_i c_j}$ is the covariance function between 
$c_i$ and $c_j$~\cite{Errors-Taylor}.
The narrow bands presented in the following figures are 
the consequence of the inclusion of the cross terms.

\begin{table}[t]
\caption{
Quality of the parametrizations measured 
in terms of chi-square  per degree of freedom ($G_{En}$, $G_E$ and $G_C$).
$\chi^2 ({\rm tot})$ is the total chi-square per degree of freedom. 
The results between brackets indicate the values 
associated with the sets not included in the fits.
}
\label{tabChi2a}
\begin{center}
\begin{tabular}{l| c c c c c}
\hline
\hline
& $\chi^2 (G_{En})$ &  $\chi^2 (G_{E})$ &  $\chi^2 (G_C)$  & $\chi^2({\rm tot})$ \\
\hline
Fit $G_{En}$   & 0.66 & (2.70) & (26.5) &   \\
Global fit (mc) &  2.65 & 2.03  & 6.73 & 2.94 \\
{\bf Global fit} & 0.77  & 0.67 & 1.15 &  0.77 \\
Fit $G_{En}$, $G_E$     &  0.66 & 0.63 & (2.33) &  \\
\hline
\hline
\end{tabular}
\end{center}
\end{table}

One can now comment on the results of the fits.
In fig.~\ref{figGEn-v1}, we can notice 
that the direct fit to the data 
provide a much better description 
of the $G_{En}$ data (reduced chi-square 0.66)
than the global fit restricted 
to the meson cloud term (reduced chi-square 2.65).
In the last case the fit 
overestimates the $G_{En}$ data 
for $Q^2 > 0.25$ GeV$^2$.
It is worth noticing that, 
although the combined fit reduces the errors 
associated with the parameters, and 
consequently the width of the bands associated with $G_{En}$,
that does not imply that the data 
description is more accurate,
as can be inferred from the reduced chi-square values.

As for the quadrupole form factors 
displayed in fig.~\ref{figGEGC-v1},
we can conclude that the combined fit (at blue) 
improves the description of the quadrupole 
form factor data compared to a fit 
that excludes those data (at red).
The improvement is more significant for $G_C$.
Both fits show a reasonable agreement with the data near $Q^2=0$.
Focusing on the global fit,
it is important to mention that 
the results are the consequence of the 
overestimated results for $G_{En}$, which enhance 
the values of $G_E$ and $G_C$ at low $Q^2$ in about 30\%.
We can also note that $G_E$ is poorly described for $Q^2 > 0.4$ GeV$^2$
and that the global fit of $G_C$ underestimates
the data in the region $Q^2 \approx 0.2$ GeV$^2$.

To summarize this first analysis, one can say that a global fit 
which neglects the valence quark contributions appears
to give a reasonable description of the quadrupole form factor data at low $Q^2$. 
Those results,  however,
are a  consequence of a rough description of the $G_{En}$ data.


\subsection{Global fit (including the valence quark contributions)}
\label{sec-GlobalFit}

We consider now a global fit to the data
that take into account the contribution of the valence quarks
on the quadrupole form factors $G_E$ and $G_C$.
The inclusion of the valence quark term 
introduces additional uncertainties in the  
calculations, since there are now uncertainties 
associated with the quark model estimates 
of the functions $G_a^{\rm B}$ ($a=E,C$).

The simplest way of taking into account 
those uncertainties in the fit it is to modify 
the experimental standard deviation
$\sigma_i$ associated with $G_a$ according to 
$\sigma_i^2 \to \sigma_i^2 + \sigma^2 (G_a^{\rm B})$,
in the chi-square calculation,
where $\sigma (G_a^{\rm B})$ is the theoretical error
associated with the result for $G_a^{\rm B}$.
The previous modification is justified 
in the calculation of the factor $G_a -(G_a^{\rm B} + G_a^\pi)$,
when the errors are combined in quadrature.
The main consequence of the previous procedure 
is the reduction of the impact of the $G_E$ and $G_C$
data in the global fit, since the contribution 
of each term is weighted by the factor $1/\sigma_i^2$. 
In those conditions the chi-square 
associated with the experimental data differ 
from the {\it effective}  chi-square 
obtained when we include the quark model uncertainties
(the model uncertainties reduce the chi-square values).
The interpretation of the reduced chi-square values 
become less clear, then, since it is not 
restricted exclusively to the experimental data.

For the reasons mentioned above, we choose 
not to take into account explicitly the 
uncertainties associated with the quark model 
in the global fit to the data.
Alternatively, we determined the chi-square  
taking into account the experimental errors and include 
the quark model uncertainties afterward in 
the representation of the results for $G_E$ and $G_C$.
With this procedure, we preserve the usual interpretation of  
reduced chi-square based on the experimental data.
The quark model uncertainties are taken into account later
in the discussion of the results for 
the quadrupole form factors.

The parameters associated with the best fit to the 
$G_{En}$, $G_E$ and $G_C$ data, 
including the valence quark and meson cloud
contributions on the quadrupole form factors 
are presented in the third row of table~\ref{tabMod0}.
The results for the reduced chi-square 
associated with the different sets of data 
are presented in the third row of table~\ref{tabChi2a}.

The chi-square values obtained in the global fit 
seems to suggest that all the data subsets ($G_{En}$, $G_E$ and $G_C$) 
are well described, since the reduced chi-square is smaller than the unit, 
or just a bit larger, in the case of $G_C$.
One can then conclude that if the theoretical uncertainties are small,
the inclusion of the valence quark contributions
improves the global description of the data.

\begin{figure}[t]
\vspace{.6cm}
\centerline{\mbox{
\includegraphics[width=3.0in]{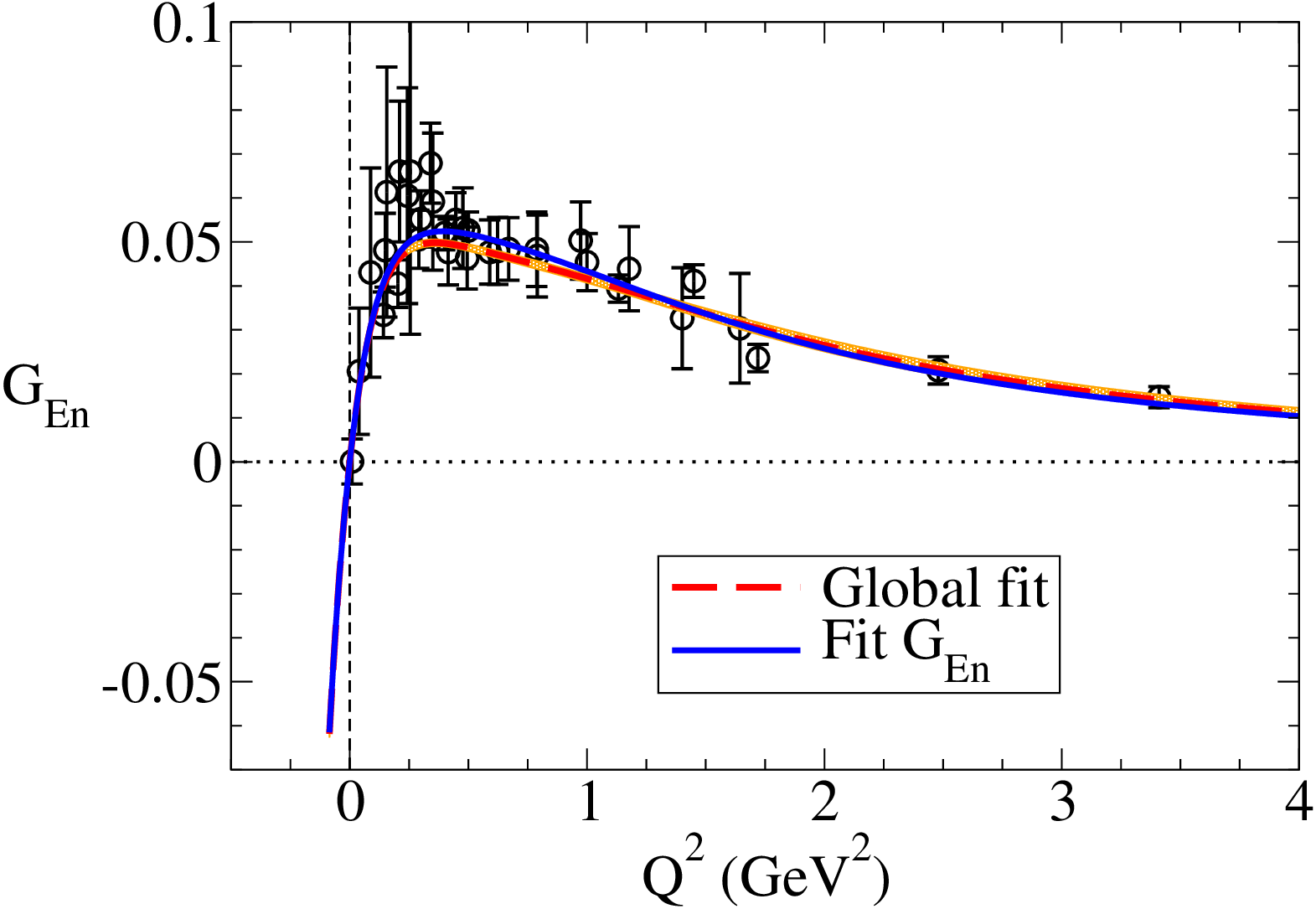}}}
\caption{\footnotesize
Neutron electric form factor. 
Best fit to the data (band and dashed line)
compared with the direct fit and with the $G_{En}$ data (solid line). 
Data from~\cite{Gentile11,Schiavilla01}.
}
\label{figGEn-v1a}
\end{figure}
\begin{figure}[t]
\vspace{.6cm}
\centerline{\mbox{
\includegraphics[width=3.0in]{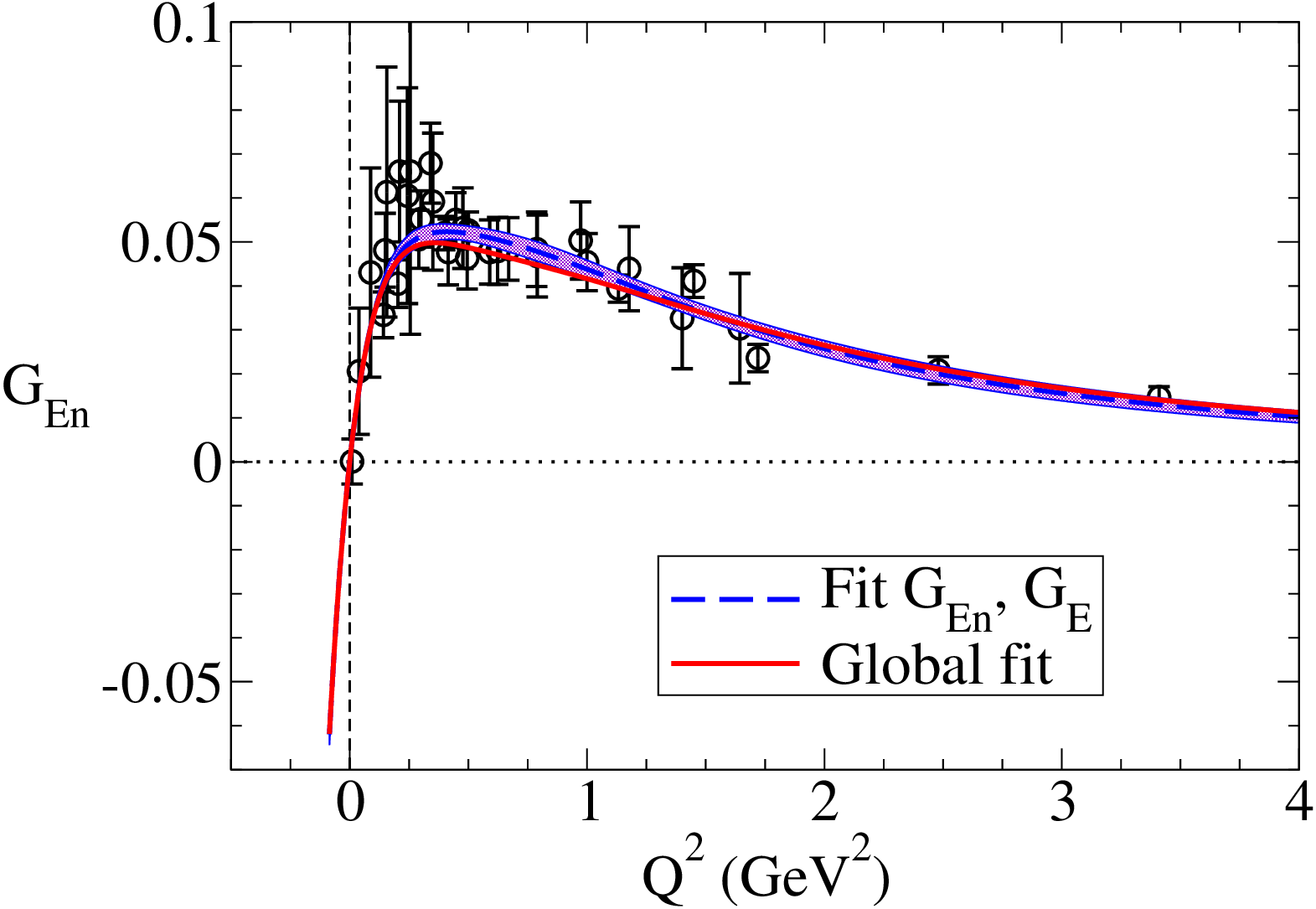}}}
\caption{\footnotesize
Neutron electric form factor. Best fit to the data (solid line)
compared with the best fit to the $G_{En}$ and $G_E$ data (excluding the $G_C$ data).
The bands represent the errors associated with 
the partial fit ($G_{En}$ and $G_E$).
The errors associated with the global fit are not included for clarity 
(see fig.~\ref{figGEn-v1}).
Data from~\cite{Gentile11,Schiavilla01}.}
\label{figGEn-v2}
\end{figure}

The numerical results for 
$G_{En}$ and for  the quadrupole form factors $G_E$ and $G_C$
are discussed in the following subsections,
including the uncertainties associated with the quark model from~\cite{LatticeD}.
The relative uncertainties associated with the quark model 
can be determined numerically 
combining the lattice QCD results $G_a^{\rm latt}$ 
with the model estimate of that result in lattice $G_{a}^{\rm B}$.
One conclude then that the errors associated with the functions 
$G_E^{\rm B}$ and $G_C^{\rm B}$ can be expressed\footnote{The 
relative uncertainty of the model for $G_a^{\rm B}$
is determined by $\frac{\sigma (G_a^{\rm B}}{G_a^{\rm B}}$.  
We can estimate 
$\left(\frac{ \sigma (G_a^{\rm B})}{G_a^{\rm B}} \right)^2$ 
using the average $\left<\left( \frac{G_a^{\rm latt} -G_a^{\rm B}}{G_a^{\rm B}} 
\right)^2 \right>$. 
In the calculation the weight of each point 
is $1/\sigma^2 (G_a^{\rm latt})$.}
as $\sigma (G_E^{\rm B}) =0.45 \,G_E^{\rm B}$  and 
$\sigma (G_C^{\rm B}) =0.33\, G_C^{\rm B}$. 
The comparison of the results of the quark model 
with lattice QCD results is possible because 
the meson cloud effects are suppressed in lattice QCD simulations 
with large pion masses~\cite{LatticeD}.
The lattice QCD results for $G_E$ and $G_C$ are 
calculated from the results for $G_E/G_M$ and $G_C/G_M$,
which have uncertainties much larger than $G_M$.
In total one has 21 points for $G_E$ 
and 21 points for $G_C$ for 
$m_\pi= 411$, 490 and 593 MeV~\cite{LatticeD,Alexandrou08}.

\subsection{Neutron electric form factor ($G_{En}$)}
\label{secGEn2}

We now discuss the numerical results for $G_{En}$ presented in 
fig.~\ref{figGEn-v1a} (solid line) in comparison 
with the data from~\cite{Gentile11,Schiavilla01}. 

The quality of the description of the $G_{En}$ data 
and the narrow band of variation can be observed 
in detail in fig.~\ref{figGEn-v1a},
where we include also the results 
of the single fit to the $G_{En}$ data, 
where we neglect the effect if the $G_E$ and $G_C$ data.
In the last case, we omit the band of variation 
associated with the first model (fit $G_{En}$) for clarity.
The comparison with the results from fig.~\ref{figGEn-v1}  
is sufficient to show that the errors associated 
with the function $G_{En}$ are reduced when we 
consider the global fit.
Based on the results presented in the third row of table~\ref{tabMod0},
one can conclude that the errors associated the parameters 
decrease compared to the previous fits, justifying the thin band of variation
presented in the figure.


\begin{figure*}[t]
\vspace{.6cm}
\centerline{
\mbox{\includegraphics[width=3.in]{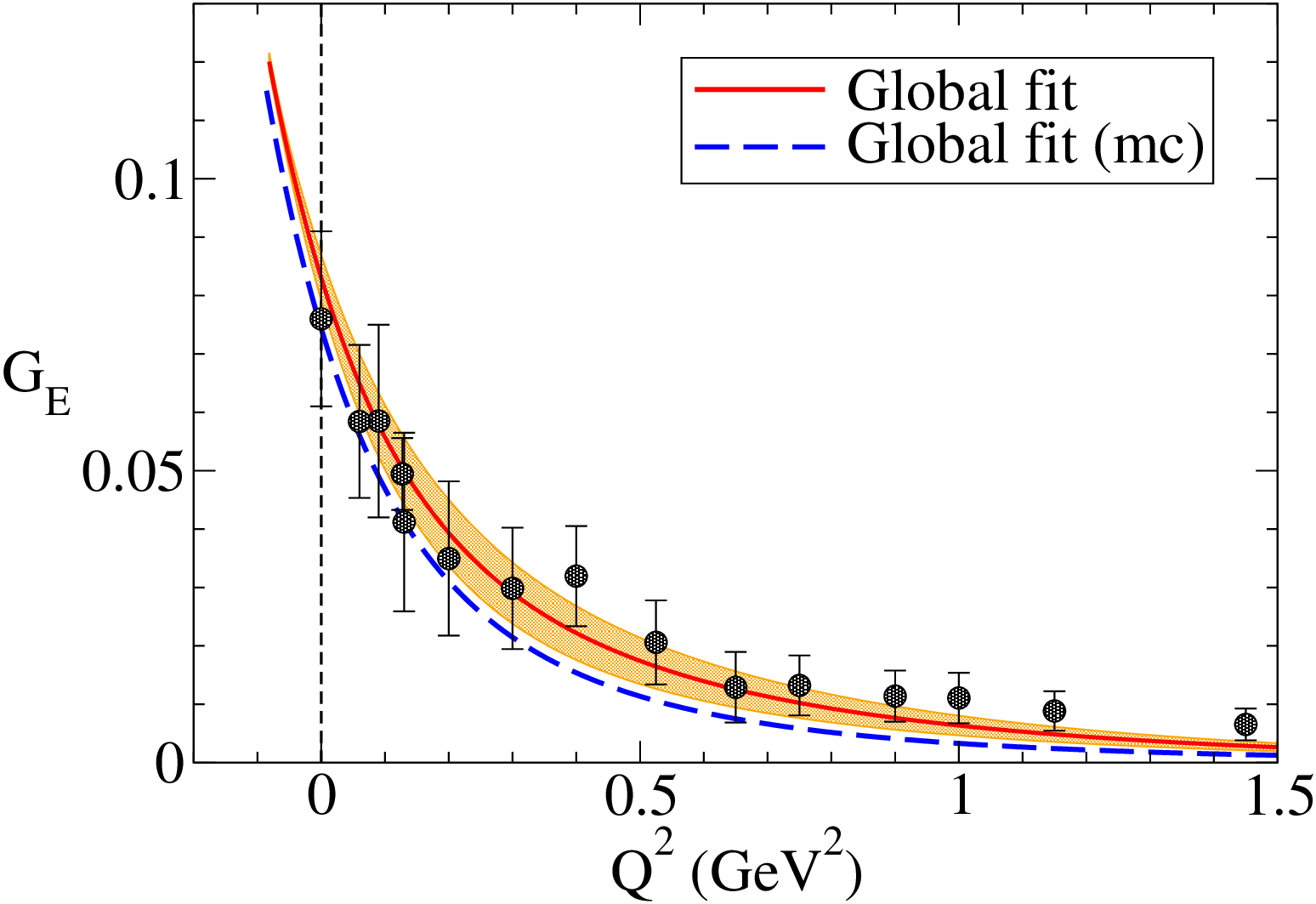}} \hspace{.8cm}
\mbox{\includegraphics[width=3.in]{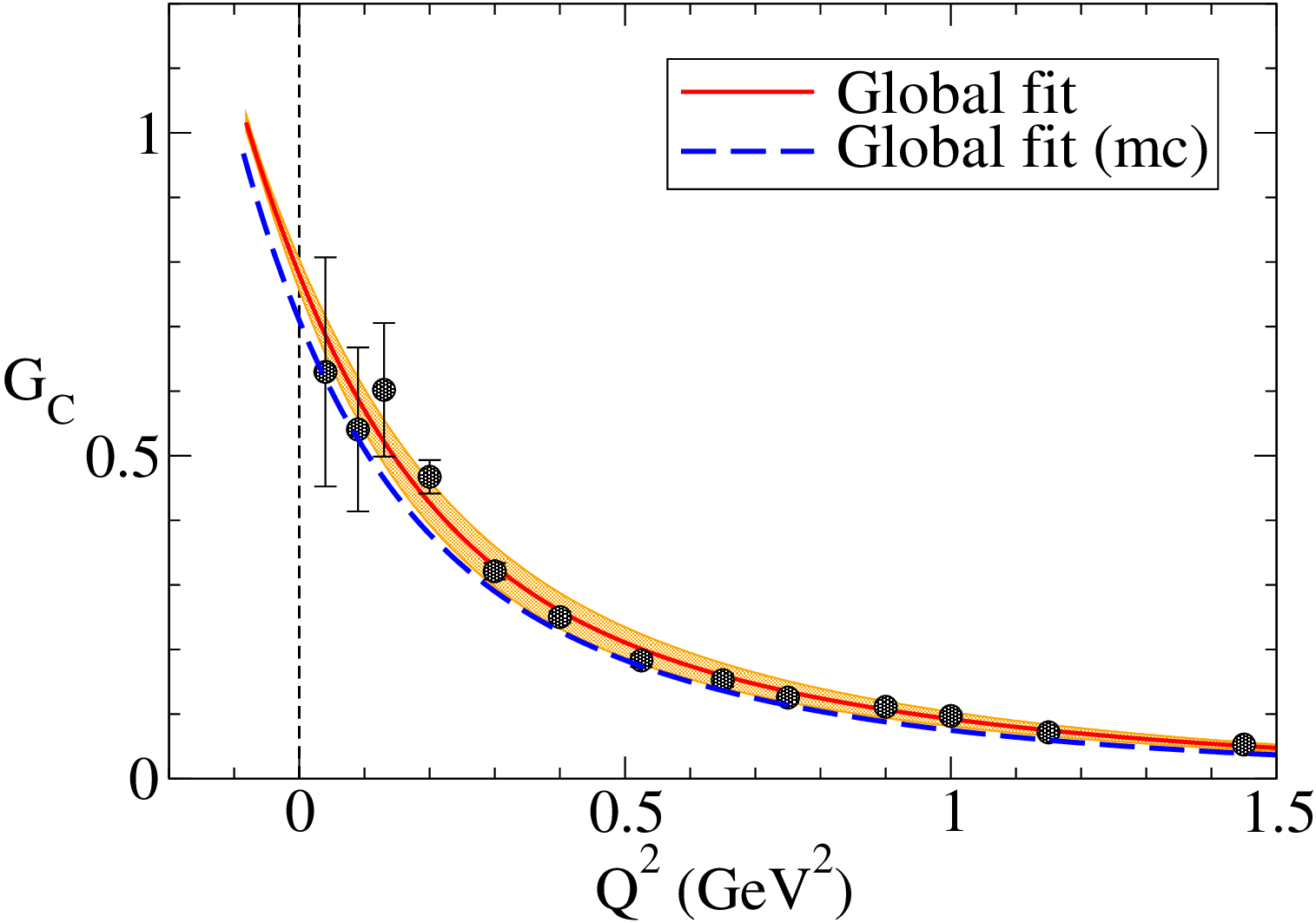}}}
\caption{\footnotesize
$\gamma^\ast N \to \Delta(1232)$ electric and
Coulomb quadrupole form factors.
Best global fit to the data (solid line, orange band).
The result of the best fit without 
valence quark contributions is also presented (dashed line, 
``mc'' stands for meson cloud) for comparison.
Data for proton targets 
from~\cite{Blomberg16a,Stave08,PDG,MIT_data,Jlab_data1,Jlab_data2}.}
\label{figGEGC-v2}
\end{figure*}

The parametrization presented in fig.~\ref{figGEn-v1a}
correspond to the value
\ba
r_n^4 = ( -0.383 \pm 0.012) \; \mbox{fm}^4.
\label{eqrn4}
\ea
This estimate is close to other estimates 
presented in the literature based on 
the Galster parametrization~\cite{Grabmayr01}
and similar parametrizations~\cite{Kaskulov04}.
For the purpose of the discussion,
we note that a Galster parametrization
that provides a good description of the  $G_{En}$ data,
associated with  $d =2.8$
correspond to $r_n^4 \simeq -0.30$ fm$^4$~\cite{Siegert-ND,Letter,Buchmann04}.

It is also worth mentioning that in a parametrization
that include the factor $G_D$, a significant contribution to $r_n^4$ 
comes from a term on $1/\Lambda^2$,
where $\Lambda^2 = 0.71$ GeV$^2$.
When we consider a Galster parametrization, 
the last contribution is about  $ -0.25$ fm$^4$.
It is then interesting to conclude that we obtain a result 
close to the Galster parametrizations, 
without including the factor $G_D$.

The comparison between the results for the third moment 
is even more compelling.
We obtain almost the same result for $r_n^6$:
$r_n^6 = (- 1.31 \pm 0.12)$ fm$^6$, using the parametri-zation 
with rational functions and $r_n^6  \simeq - 1.34$ fm$^6$, 
using the Galster parametrization with 
$d =2.8$~\cite{Siegert-ND,Letter,Buchmann04}.
We conclude then that the two paramerizations are distinguished near $Q^2=0$,
within the uncertainties, basically 
by the value of the second moment ($r_n^4$).

In figs.~\ref{figGEn-v1} and \ref{figGEn-v1a},
we can notice that the different parame-trizations
have a very similar extension to the timelike 
region, including the pseudothreshold.
Very similar results can also be obtained 
with the Galster parametrization.
This effect is explained by the dominance 
of the first two terms (first and second moments) 
in the expansion of $G_{En}$ from (\ref{eqGEn-expan}) 
near the pseudothreshold.

In order to test the dependence of the fit on the data subsets, 
we perform also  partial fits to the 
combination of the data $(G_{En}, G_E)$ and $(G_{En}, G_C)$.
This way, we test which quadrupole set, $G_E$ or $G_C$, 
has more impact in the global fit, 
and we can also infer if those sets are compatible with the $G_{En}$ data.
The result of the fit to the sets $(G_{En}, G_C)$ 
is represented in fig.~\ref{figGEn-v2} by the blue band.
The global fit ($G_{En}$, $G_E$ and $G_C$) 
is represented by the solid line.
The result of the combined fit for  $(G_{En}, G_C)$
is almost undistinguished from the global fit (solid line), 
and it is not presented for clarity.
The parameters associated with the partial fit  $(G_{En}, G_E)$ 
are presented in the last row of table~\ref{tabMod0}.
Compared to the global fit, we notice 
the increasing errors associated with the parameters. 
The corresponding values for the reduced chi-square 
are presented in the last row of table~\ref{tabChi2a}. 
From the result for $\chi^2(G_C)$ we can confirm 
that the quality of the description of the 
$G_{En}$ and $G_E$ is obtained at the expenses 
of a poorer description of the $G_C$ data.


In the literature, there is some debate 
whether there is a bump in the function $G_{En}$ 
near $Q^2=0.2$--0.3 GeV$^2$, or not~\cite{Pascalutsa07a,Friedrich03}.
We conclude that the present accuracy of the data 
is insufficient for more definitive conclusions.
Our best fit has a smooth behavior in the region under discussion.
Nevertheless, larger values of $k$ can in principle 
generate a low-$Q^2$ bump, 
once one has more high $Q^2$ data to constrain the higher order coefficients.
 
When we take into account the theoretical uncertainties 
associated with the quark model the results 
for $G_{En}$ are only somewhat modified.
The magnitude of $G_{En}$ is slightly enhanced near $Q^2= 0.5$ GeV$^2$.

\subsection{Quadrupole form factors ($G_E$, $G_C$)}
\label{secQ-Delta}

The results of the best global fit  to the 
$\gamma^\ast N \to \Delta(1232)$ quadrupole form factors (solid line)
including the theoretical uncertainties (orange band) 
are presented in fig.~\ref{figGEGC-v2},
in comparison with the respective data.
For comparison, we include also the previous fit 
where we ignore the valence quark contributions (dashed line).

Focusing first in the central values  (solid line) 
we observe an excellent agreement  
with the data in the range $Q^2=0$--1.5 GeV$^2$, for both form factors.
As mentioned before, the quality of the description is  
corroborated by the reduced chi-square values presented 
in table~\ref{tabChi2a}, 0.67 and 1.15 for $G_E$ and $G_C$, respectively.
For this agreement contributes 
the larger error bars for the data below 0.2 GeV$^2$.
The precision of the $Q^2< 0.2$ GeV$^2$ data is comparable to 
the precision of most of the neutron electric form factor data. 
We note, however, that in the present case 
it is not sufficient to look for the chi-square values,
because there are also uncertainties associated 
with the valence quark contributions.

The uncertainties displayed in fig.~\ref{figGEGC-v2}
are the combination of the pion cloud uncertainties
associated with the $G_{En}$ parametrization and the 
uncertainties associated with the quark model.
The main contribution comes from the quark model.
The magnitude of the pion cloud uncertainties
can be inferred from fig.~\ref{figGEGC-v1}.
The uncertainties are very small at pseudothreshold
because according to our estimate, the 
uncertainties associated with the valence quark contribution vanish 
and only the pion cloud component contributes 
(recall that $G_E^{\rm B}$ and $G_C^{\rm B}$ vanish at pseudothreshold).

The results from fig.~\ref{figGEGC-v2}
shows that there is a range of variation of $G_E^{\rm B}$ and $G_C^{\rm B}$
which give a better solution than 
models with no valence quark contributions.
In the figure, we can notice that 
the model with no valence quark contributions (dashed line) 
is a bit below the lower limit of the estimate 
with valence quark contributions.
One can then conclude that the inclusion 
of the bare contribution improves the description of the data.

The lower limit of the results for $G_E$ is 
obtained for a bare contribution of about $0.55\,G_E^B$,
where $G_E^B$ is the estimate of the 
covariant spectator quark model.
In this case, we cannot conclude much, 
since theoretical and experimental uncertainties 
are both large and compatible.

The improvement is more significant in the case 
of $G_C$ with a much narrower band of variation.
In this case, the model uncertainties are 
about 33\% of the model estimate $G_C^B$.
One can notice in this case, 
that although the fit with no valence quark component (dashed line)
is just a bit below the lower limit of the model estimate,
the reduced chi-square values are very different (see table~\ref{tabChi2a}).
The difference between those values is a consequence 
of the small error bars associated with the $Q^2 > 0.2$ GeV$^2$ data for $G_C$.
One concludes, then that $G_C$ is more sensitive 
to the inclusion of the valence quark component.

\begin{figure}[t]
\vspace{.8cm}
\centerline{
\mbox{\includegraphics[width=3.0in]{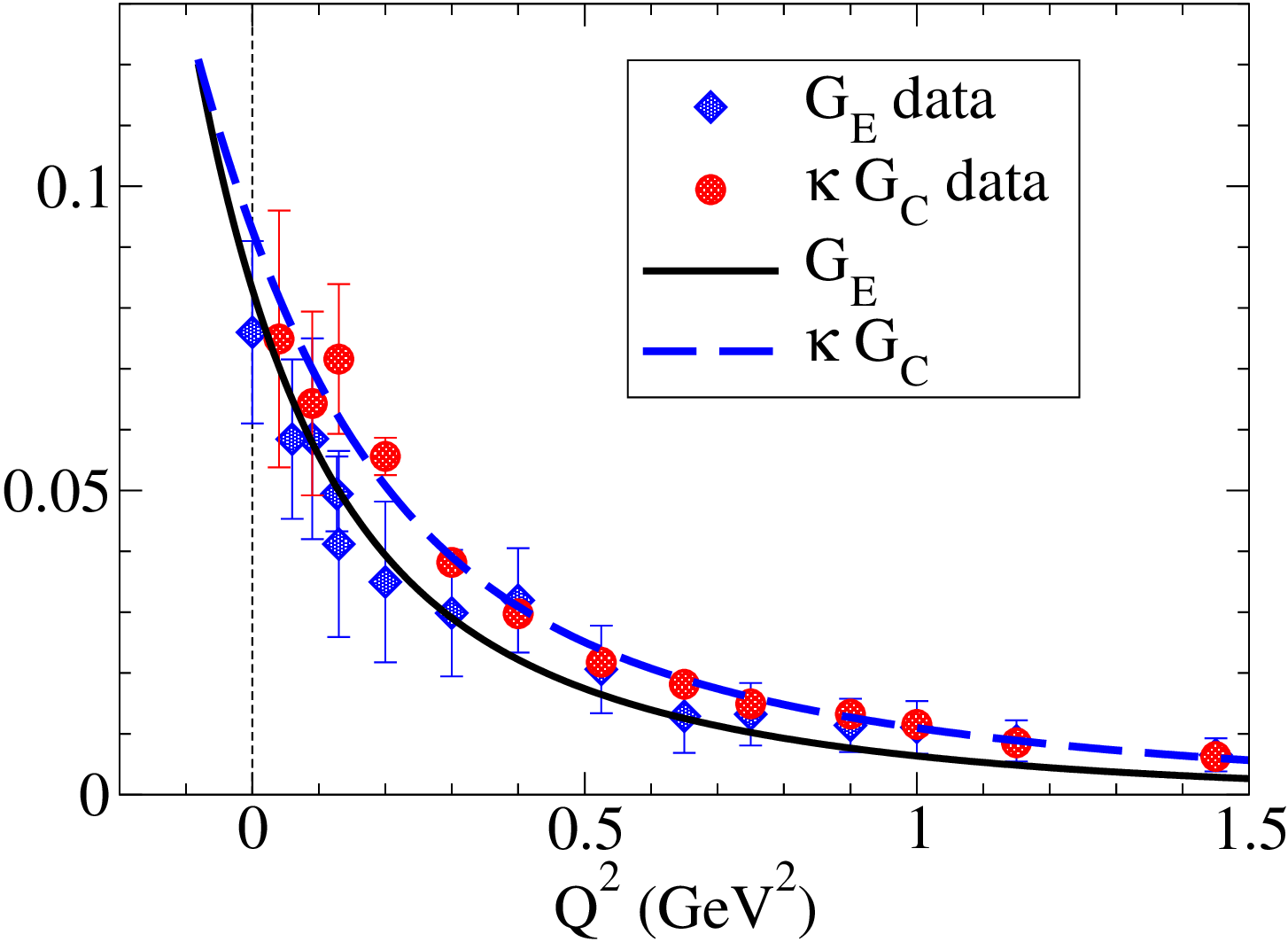}} }
\caption{\footnotesize
Results for the form factors $G_E$ and $G_C$.
Recall that $\kappa = \sfrac{M_\Delta-M}{2M_\Delta} \simeq 0.12$.
Data for proton targets 
from~\cite{Blomberg16a,Stave08,PDG,MIT_data,Jlab_data1,Jlab_data2}.}
\label{figGEGC}
\end{figure}

The comparison of results of $G_E$ with $G_C$ corrected 
by the factor $\kappa$ is presented in fig.~\ref{figGEGC}.
For clarity, we omit the uncertainties
associated with both form factors.
In this representation, it is more explicit 
the connection between the two quadrupole form 
factors and the implications of Siegert's theorem.
The convergence of the results at the pseudothreshold 
($Q^2 \simeq - 0.09$ GeV$^2$),
the consequence of Siegert's theorem,
is clearly displayed in the graph.

In comparison with the results from~\cite{Letter}, 
where $G_{En}$ is described by a Galster parametrization ($d =2.8$), 
we obtain slightly larger values for the 
quadrupole form factors at the pseudothreshold.
 
Based on the results of $G_E$ and $G_C$ at $Q^2=0$, 
one can test the accuracy of the large $N_c$ estimate 
$G_E = \sfrac{M_\Delta^2-M^2}{4M_\Delta^2}G_C$, 
apart ${\cal O}(\sfrac{1}{N_c^2})$ relative corrections~\cite{Pascalutsa07a}.
The previous relation is equivalent to the identity 
$R_{EM}(0) = R_{SM}(0)$~\cite{Letter,Pascalutsa07a}.
According to the results 
from figs.~\ref{figGEGC-v2} and \ref{figGEGC}, one obtains  $G_E(0) = 0.083 \pm 0.003$ 
and $\sfrac{M_\Delta^2-M^2}{4M_\Delta^2}G_C(0) = 0.082 \pm 0.004$,
in agreement with the large $N_c$ prediction,
within the uncertainties.

\subsection{Square radii $r_E^2$ and $r_C^2$}
\label{secRadii}

The difference of behavior between the two 
quadrupole form factors can be better understood when we look 
at the square radius associated 
with the quadrupole form factors $G_E$ and $G_C$,
defined by~\cite{Buchmann09a,SiegertD,Forest66}
\ba
r_a^2 = -\frac{14}{G_a(0)}
\left. \frac{d G_a}{d Q^2} 
\right|_{Q^2=0}
\label{eqRadius}
\ea
where $a=E,C$.
The numerical results for $r_E^2$ and $r_C^2$ for 
the parametrizations discussed here are 
presented in table~\ref{tabRadius}.

If we ignore the effect of the valence quarks,  
one can calculate $r_E^2$ and $r_C^2$ very easily 
using eq.~(\ref{eqGE1})-(\ref{eqGC1}). 
One obtains then 
$r_E^2 = \frac{7}{10} \frac{r_n^4}{r_n^2} + \sfrac{7}{M_\Delta (M_\Delta -M)}$
and $r_C^2 = \frac{7}{10} \frac{r_n^4}{r_n^2}$.
We conclude then that $r_E^2 -r_C^2$ is of order $N_c^0$,
and it is not suppressed in the large $N_c$ limit.
The relation for  $r_C^2$ was derived previously in~\cite{Buchmann09a}, 
assuming the dominance of the pion cloud contribution.
Using the result (\ref{eqrn4}) for $r_n^4$,
one obtains the estimates 
$r_E^2 = 3.06 \pm 0.07$ fm$^2$ and 
$r_C^2 = 2.31 \pm 0.07$ fm$^2$.
The uncertainties in the previous results 
are the consequence of the results for $r_n^4$,
affected by the errors associated 
with the parametrization $G_{En}$.
Both estimates are about 10\% larger than the results 
for the pion cloud contribution presented in table~\ref{tabRadius},
and are therefore consistent with a 10\% 
correction for both form factors  near $Q^2=0$,
due to the inclusion of the valence quark contributions,
as discussed in sect.~\ref{secPC}.

Combining the analytic expressions described previously 
for $r_E^2$ and $r_C^2$, 
one has $r_E^2 - r_C^2 =  \sfrac{7}{M_\Delta (M_\Delta -M)} \simeq 0.75$ fm$^2$.
Note that this result is independent 
of the general form used for the function $G_{En}$,
and therefore independent of $r_n^4$.

The previous estimate differs from the result presented 
in table~\ref{tabRadius},   
mostly because of the valence quark effects are not taken into account  
as discussed above (10\% reduction).
Correcting the previous result with the valence quark effect,
assuming the 10\% correction, 
one obtains then $(r_E^2 -r_C^2)_\pi \simeq 0.68$ fm$^2$.
The subindex was included to emphasize that 
the estimate concerns the pion cloud component.
The previous estimate is still in disagreement 
with the result from table~\ref{tabRadius}, in about $-0.04$ fm$^2$.
In this discussion, we omit the uncertainties 
and focus on the results obtained with the central values
presented in the table.

To understand the small difference of $-0.04$ fm$^2$,
we need to look into the details of the calculation,
and for the very small deviations from the 10\% correction, 
assumed for the valence quark component.
If we consider deviation of $\delta_E$ and $\delta_C$ 
from the 10\% correction to the 
form factors $G_E$ and $G_C$, respectively,
we conclude that the residual correction 
can be expressed as $0.9^2 (\delta_C -\delta_E) r_{E0}^2$,
where the first factor accounts for the 
10\% correction in the normalizations, 
and $r_{E0}^2$ is the estimate of $r_E^2$ 
in a model with pion cloud dominance, mentioned above.
Combining all effects with $r_{E0}^2 \simeq 3.06$ fm$^2$, 
one obtain   $0.9^2 (\delta_C -\delta_E) r_{E0}^2 \simeq -0.04$ fm$^2$,
which explain at last the result $(r_E^2-r_C^2)_\pi \simeq 0.64$ fm$^2$.
The last correction is then the consequence of 
the product of a very small factor by a large factor ($r_{E0}^2$).

The contributions from the valence quark component 
depend on the details of the quark model, 
and are therefore model dependent.
Using the model described previously, 
we obtain $(r_E^2 -r_C^2)_{\rm B} \simeq - 0.06$ fm$^2$,
where B labels the bare contribution.
Combining the two contributions, 
we obtain $r_E^2 -r_C^2 \simeq  0.58$ fm$^2$,
as presented in table~\ref{tabRadius}.
When we take into account all 
the errors we obtain $r_E^2 -r_C^2 =  0.58 \pm 0.23$ fm$^2$ 
(see table~\ref{tabRadius}).

\begin{table}[t]
\caption{Contributions to the 
quadrupole square radius (bare and pion cloud): $r_E^2$, $r_C^2$ 
and difference $r_E^2- r_C^2$. 
The results between brackets indicate the 
uncertainties.}
\label{tabRadius}
\begin{center}
\begin{tabular}{c | c c c}
\hline
\hline
 & $r_E^2$ (fm$^2$) & $r_C^2$ (fm$^2$)  & $r_E^2- r_C^2$ (fm$^2$)  \\
\hline
B      & $-0.410$     &  $ -0.347$  & $-0.063$\\
       & $\;\;\;(0.169)$     &   $\;\;\; (0.107)$   &  $\;\;\; (0.200)$  \\ 
$\pi$  & $\;\;\; 2.742$   &  $\;\;\;2.100$     & $\;\;\; 0.643$ \\
       & $\;\;\;(0.080)$  &  $\;\;\;(0.064)$   & $\;\;\; (0.102)$  \\
\hline
Total  & $\;\;\;2.331$    &  $\;\;\;1.752$      &   $\;\;\; 0.580$ \\
       & $\;\;\;(0.187)$  &  $\;\;\;(0.134)$    &   $\;\;\; (0.225)$   \\
\hline
\hline
\end{tabular}
\end{center}
\end{table}

Smaller values for $r_C^2$ have been suggested by empirical parametrizations 
of the quadrupole form factor data 
consistent with Siegert's theorem~\cite{Siegert-ND,SiegertD}.
We note, however, that those estimates 
are strongly affected by the 
$Q^2=0$--0.15 GeV$^2$ data for $G_C$, from~\cite{Stave08,MIT_data,Sparveris13}.
Therefore, the conclusions based on those data have to be reviewed 
in light of the new data from~\cite{Blomberg16a}.

An important conclusion of the previous discussion 
is that in order to obtain a reliable estimate of $r_E^2$ and $r_C^2$
from the $G_E$ and $G_C$ data, 
we need more accurate measurements from $G_E$ and $G_C$ near $Q^2=0$.

Overall, we can conclude that 
the large values obtained for $r_E^2$ and $r_C^2$
are the outcome of the pion cloud dominance.
The magnitudes of $r_E^2$ and $r_C^2$ around 2 fm$^2$
can be interpreted physically,  
as the result of the increment of the size of the constituent quarks 
due to the $q \bar q$ pair/pion cloud dressing~\cite{Buchmann09a,SiegertD}.
More specifically, those magnitudes can be understood looking 
at the pion Compton scattering wavelength, 
$r_\pi \approx 1/m_\pi$, 
which characterize the pion distribution inside the nucleon~\cite{Buchmann09a}.
Numerically, the result $r_E^2 \approx r_C^2 \approx 2$ fm$^2$ is then 
the consequence of $r_E^2 \approx r_C^2 \approx r_\pi^2$.
A more detailed discussion on this subject can be found in~\cite{Buchmann09a}.

Concerning the difference 
$r_E^2 - r_C^2 = 0.6 \pm 0.2$ fm$^2$,
one can conclude that it is mainly a consequence of 
Siegert's theorem, since the difference between $r_E^2$ 
and $r_C^2$ is the consequence of the extra factor 
included in the pion cloud parametrization $G_E^\pi$, 
in order to satisfy Siegert's theorem.


\subsection{General discussion}

Once presented our final results for the 
neutron electric form factor and for 
$\gamma^\ast N \to \Delta(1232)$ quadrupole form factors,
one can discuss some theoretical 
and experimental aspects related to the present results.

For the good agreement between the model
and the data for the quadrupole form factors,
displayed in figs.~\ref{figGEGC-v2} and \ref{figGEGC}
contributes the valence quark component
estimated with a covariant quark model.
We stress that this part of the calculation is model dependent
and it is then limited by some theoretical uncertainties.
Those uncertainties are larger in the case 
of the electric form factor $G_E$.
Alternative estimates of the valence quark 
contribution, with the same sign, can also improve the description 
of the quadrupole form factor data.
An example is the bare para-metrization 
of the Sato-Lee model in the $Q^2 > 0$ region~\cite{LatticeD,SatoLee}.
Another example is the estimates based on the Dyson-Schwinger formalism,
which are expected to be closer to the upper 
limit of the present calculation.

A test to the valence quark contributions can 
be performed in a near future with lattice QCD simulations 
in the range of $m_\pi=0.2$--0.3 GeV~\cite{Alexandrou11}, 
a region close the physical point, but 
where hopefully the pion cloud contamination is small.
The increase of the number of lattice QCD simulations 
and of the accuracy of those simulations 
can also help to improve the accuracy of the present quark model estimates.

Another relevant point of discussion is the accuracy of the data.
The data associated with $G_{En}$ 
have in general large error bars, 
which difficult the derivation of an accurate parametrization.
The $G_E$ and $G_C$ data below 0.15 GeV$^2$  
are also affected by large error bars.
In that region the experimental uncertainties 
are larger than the quark model uncertainties.
For all those reasons it is not hard to find parametrizations 
that provides a good description of the overall data, 
including the low-$Q^2$ region (small chi-square).
The inclusion of the $G_E$ and $G_C$ data provide,
however, additional constraints to the function $G_{En}$,
which help to pin down the shape of $G_{En}$.

Another important aspect of the present work 
is the sensitivity of the fits to the data,
particularly in the  region $Q^2=0$--0.2 GeV$^2$.
This effect can be illustrated by the realization that 
the trend of the function $G_C$ 
changed with the more recent data~\cite{Blomberg16a}.
A fit that includes the $G_C$ data from~\cite{Stave08,MIT_data}
cannot describe the data with the same accuracy as 
a fit that include the more recent data~\cite{Siegert-ND}.
In conclusion, new data for $G_C$ have also 
a significant impact on the solution for $G_{En}$.

To finish the present discussion, 
we note that we can also test the results of the functions $G_E$ and $G_C$
combining lattice QCD results 
with estimates based on expansions on the pion mass, 
derived from effective field theories~\cite{Pascalutsa05,Gail06,Hilt18}.
In these conditions the consistency between the results from lattice QCD
and the experimental data can be checked,  
and therefore, the consistency between QCD and the real world.

\section{Outlook and conclusions}
\label{secConclusions}

In the present work, we derive 
a global parametrization of the neutron electric form factor 
and the $\gamma^\ast N \to \Delta(1232)$ quadrupole form factors, $G_E$ and $G_C$.
To relate the pion cloud contribution to $G_E$ and $G_C$ with 
the neutron electric form factor 
we use improved relations  derived in the large $N_c$ limit 
in order to verify Siegert's theorem exactly.

The success of the global parametrization 
of $G_{En}$,  $G_E$ and $G_C$ is an indication of the importance   
of the pion cloud in the neutron and 
in the $\gamma^\ast N \to \Delta(1232)$ transition.
This correlation is suggested by 
the large $N_c$ limit, by the $SU(6)$ symmetry breaking,
and by calculations based on non relativistic constituent quark models.

For the agreement between the model calculations and the empirical data for 
the $\gamma^\ast N \to \Delta(1232)$ quadrupole form factors also
contribute the small valence quark components estimated  by a covariant quark model,
calibrated previously by the results of lattice QCD simulations.
Although limited by some model uncertainties,
the estimates of the valence quark contributions
improve the description 
of the data compared to models with no bare contributions.
Estimates of the valence quark effects based on 
other frameworks may also improve the description 
of the $\gamma^\ast N \to \Delta(1232)$ quadrupole form factors.

The global fit of the  $G_{En}$,  $G_E$ and $G_C$ data,
based on rational functions, 
show that the overall data, and the $G_{En}$ data, in particular, 
is compatible with a smooth description of the neutron electric form factor
with no pronounced bump at low $Q^2$. 
The best description of the data 
is obtained when we consider a parametrization
of $G_{En}$ associated with $r_n^4 = -0.38 \pm 0.01$ fm$^4$.

We conclude that the square radii associated with 
the quadrupole form factors $G_E$ and $G_C$ are large, 
as a consequence of the pion cloud effects 
(long extension of the pion cloud).
We also conclude  that the square radii, 
$r_E^2$ and $r_C^2$ are constrained by the relation 
$r_E^2- r_C^2 = 0.6 \pm 0.2$ fm$^2$.
The previous relation is a consequence of Siegert's theorem and 
of the dominance of the pion cloud contributions 
on the  quadrupole form factors $G_E$ and $G_C$.

The present parametrization of the neutron electric form factor
is still derived from data with significant error bars below 0.2 GeV$^2$.
Future experiments with more accurate data for $G_{En}$, 
and the $\gamma^\ast N \to \Delta(1232)$ quadrupole form factors 
can help to elucidate the shape of the function $G_{En}$ at low $Q^2$.
Of particular interest are the upcoming results 
of the JLab 12-GeV upgrade~\cite{NSTAR,NSTAR2017}.

\section*{Acknowledgments}
The author thanks Mauro Giannini 
for helpful discussions.
This work was supported by the Funda\c{c}\~ao de Amparo \`a 
Pesquisa do Estado de S\~ao Paulo (FAPESP):
project no.~2017/02684-5, grant no.~2017/17020-BCO-JP. 


\end{document}